\documentclass[twocolumn,showpacs,superscriptaddress,nofootinbib,prd,floatfix]{revtex4}
\usepackage{longtable}
\usepackage{subfigure}
\usepackage{color}
\usepackage{graphicx}
\usepackage{verbatim}
\usepackage{psfig}

\begin{document}
\title{Reconstruction and constraining of the jerk parameter from OHD and SNe Ia observations}

\author{Zhong-Xu Zhai}
\affiliation{Department of Physics, Institute of Theoretical
Physics, Beijing Normal University, Beijing, 100875, China}

\author{Ming-Jian Zhang}
\affiliation{Department of Physics, Institute of Theoretical
Physics, Beijing Normal University, Beijing, 100875, China}

\author{Zhi-Song Zhang}
\affiliation{Department of Aerospace Engineering, School of
Astronautics, Harbin Institute of Technology (HIT), Harbin
Heilongjiang, 150001, China}

\author{Xian-Ming Liu}
\affiliation{Department of Physics, Hubei University for
Nationalities, Enshi Hubei, 445000, China} \affiliation{Department
of Physics, Institute of Theoretical Physics, Beijing Normal
University, Beijing, 100875, China}

\author{Tong-Jie Zhang}\email{tjzhang@bnu.edu.cn}
\affiliation{Department of Astronomy, Beijing Normal University,
Beijing, 100875, China} \affiliation{Center for High Energy Physics,
Peking University, Beijing, 100871, China}

%\author{Wen-Biao Liu}
%\affiliation{Department of Physics, Institute of Theoretical Physics, Beijing Normal University, Beijing, 100875, China}

\begin{abstract}
Compared with the plentiful researches of the Hubble parameter and
deceleration factor, the third time derivative of the scale factor
$a(t)$ in the FRW cosmology, namely, the jerk parameter $j$ still
lacks attention. In order to study the properties of $j$, we propose
several kinds of parameterizations of $j(z)$ as a function of the
redshift $z$. By setting the standard $\Lambda$CDM model as the
fiducial model, we constrain the jerk models with the observational
Hubble parameter data (OHD) and Type Ia Supernovae (SNe)
observations. We find that the perturbation of $j(z)$ favors a value
of nearly zero and the $\Lambda$CDM is well accommodated by the jerk
reconstruction. We also compare the powers of OHD and SNe in
constraining the jerk models in detail, and find that the newly released OHD
measurement at $z=2.3$ can improve the constraint significantly,
even tighter than the SNe one. Furthermore, we analyze the jerk
models by calculating the Hubble parameter, equation of state, the
deceleration factor and $Om(z)$ diagnostic. Our results show that
the universe is indeed undergoing an accelerated expansion phase
following the matter-dominated one, which is consistent with the
standard model by observations.

\end{abstract}

 \maketitle

\section{Introduction} \label{sec:introduction}

One of the most important discoveries of the modern cosmology is the
accelerated expansion of the universe. This phenomenon was first
discovered by the type Ia supernovae observations
\cite{SNe..Riess,SNe..Hicken,SNe..Perlmutter}, and later further
confirmed by the measurements of the cosmic microwave background
(CMB) \cite{CMB..Spergel,CMB..Komatsu}, the baryon acoustic
oscillation (BAO) \cite{BAO..Eisenstein,BAO..Percival} and so on.
Now more than ten years later, this phenomenon has been accepted
widely. As pointed out in Ref.\cite{review..Weinberg}, the question
concerning this is no longer whether the universe is accelerating,
but why.

In order to give a reasonable explanation to this scenario, a great
variety of attempts have been done. These works include the dark
energy models which involve the introduction of exotic matter
sources, and the modified gravity models which relates to the
changes of the geometry of the spacetime
\cite{dark..energy..review..Tsujikawa,dark..energy..review..Sola,dark..energy..review..Li,dark..energy..review..Capozziello,dark..energy..review..Trodden}.
Although these models can
solve some problems and fit the observational data, they also have
their own difficulties. For example, the standard $\Lambda$CDM model
is considered to be the simplest and most natural one which shows
great consistence with the observational data\cite{LCDM..Jassal,LCDM..Wilson,LCDM..Davis,LCDM..Allen}. In this model, the
cosmological constant $\Lambda$ is considered to be the dark energy
component of the universe. However, it is also challenged by the
fine-tuning problem and the coincidence problem. So the study of
explaining the accelerated expansion is still continued and the new
models are being proposed
\cite{cons..prob..S..Weinberg,cons..pro..A. Vilenkin,cons..pro..J.
Garriga,cons..pro..G. Caldera-Cabral}.

For obtaining more information of the evolutionary behavior of the
universe, one can study the time derivative of the scale factor
$a(t)$ with respect to the redshift $z$ in the frame of a FRW
universe, such as the Hubble parameter $H=\dot{a}/a$, the
deceleration factor $q=-\ddot{a}/(aH^{2})$ and so on. As a direct
indication of the decelerated/accelerated expansion, the parameter
$q$ has been studied from both the observational and theoretical
views, including the constraints from the observational data, the
analysis of a particular model, or the reconstruction by some
statistical methods such as the Principle Component Approach (PCA)
and so on
\cite{q..Cunha,q..Virey,q..Yi,q..Otto,PCA..Huterer,q..Shapiro,q..Campo}.
As a higher order derivative of the scale factor, the jerk parameter
$j=-\dddot{a}/(aH^{3})$ is related to the third time derivative of
$a$ (we notice that in some earlier works, there is no negative sign
in the definition as Ref.\cite{j..Blandford,j..Visser} and so forth.
However, some literatures contain the negative sign as
Ref.\cite{j..model2..1,j..model2..2}. We point out the difference
here in order to avoid the confusion). It is a measurement of the
variation of $q$ and can be used as an indication to predict the
future of the universe. Because the higher-order derivatives can
characterize the dynamics of the universe, it could be related to
the emergence of sudden future singularities
\cite{future..Dabrowski,future..Dabrowski2}. Another example is the
$\Lambda$CDM cosmology where $j_{\Lambda}=-1$ implies that the
universe will continue to expand with an acceleration because of the
cosmological constant. Except that, the jerk parameter $j$ is also
applied in the statefinder diagnostic to discriminate different dark
energy or modified gravity models
\cite{statef..Sahni,statef..Alam,j..fR,j..Dunajski,j..Poplawski}.
Although the single $j$ can not identify some similar models as
$\Lambda$CDM and Einstein de-Sitter universe, its combination with
$q$ can comprise an identification in a wider range of cosmological
models. Compared with the plentiful research works of $q$, the jerk
parameter has not been fully explored at present
\cite{j..Xu,j..Xu2,j..Lu,j..model2..1,j..model2..2,j..Rapetti,j..first}.
It is therefore natural to study the jerk parameter because of its
importance in cosmology.

Among the numerous cosmological models, the standard $\Lambda$CDM
model can fit most of the observations and is considered as the best
one\cite{LCDM..Maor,LCDM..Basilakos,dark..energy..review..Tsujikawa}.
Thus, it is reasonable to set the $\Lambda$CDM model as the fiducial
model and thus $j=-1$ is an important reference. In our calculation,
we will mainly measure the departure of $j$ from -1 by its
parameterizations. This will be a direct generalization of the
standard $\Lambda$CDM model. Specifically, we will reconstruct jerk
as $j(z)=-1+departure$ and measure the departure term with the
observational data. The results can give us the impression if the
universe in the past was strictly the standard model or not.

In our work, the constraints on $j$ are presented by the use of the
latest Union 2.1 supernovae data (SNe)
\cite{Union2.1..Suzuki} and the observational
Hubble parameter data (OHD) \cite{OHD..Farooq}. As the two widely used measurements at
low redshift, SNe and OHD have been applied in dozens of
cosmological researches
\cite{OHD..Farooq,OHD..ChenY,SNe..Mueller,SNe..Lu,SNe..Grande,OHD..Basilakos,OHD..ZhangH,OHD..Duan,OHD..Liao,OHD..Cao,OHD..Cao2,SNe..Godlowski,OHD..Lazkoz,OHD..Wei,OHD..Wu,OHD..Kurek,SNe..ZhangX}.
The comparisons between them were also discussed deeply and widely
\cite{OHD..Ma,OHD..ZhangT,OHD..Zhai,OHD..Lin}. Therefore, we also
compare the powers of SNe and OHD in constraining the models of jerk
parameterizations and analyze the differences between them.

Our paper is organized as follows: In Sec. \ref{basic}, the basic
formulas of the kinematical models and the reconstruction of jerk
parameter are presented. In Sec.\ref{constraint}, the constraints by
use of SNe and OHD data sample are obtained and analyzed. Our
discussions and conclusions are given in Sec.\ref{conclusion}.

%then reconstruct the evolution of $j$ with
%the PCA method \cite{PCA..Huterer}. Because the PCA is a model independent method, therefore we can obtain the evolutionary information of $j$ without setting any particular models.
%Since the introduction of PCA to cosmology in 2002, it has been widely used such as the equation of state of dark energy \cite{PCA..Huterer},
%the Hubble parameter $H(z)$ \cite{PCA..Hubble}, the deceleration parameter $q$ \cite{q..Shapiro},
%the redshfit drift \cite{PCA..zdot} and so on. Therefore, it is expectable to consider the combination of PCA method and the constraints to study the jerk parameter.

\section{Kinematical models and the constraints}\label{basic}

\subsection{Reconstruction of $j(z)$} \label{sec:re_j}
Let's start with the Friedmann-Robertson-Walker (FRW) metric which
describes a homogenous and isotropic universe
\begin{equation}
  ds^{2}=-dt^{2}+a^{2}(t)\left[\frac{dr^{2}}{1-kr^{2}}+r^{2}d\Omega^{2} \right],
\end{equation}
where $k$ is the spatial curvature and for simplicity, we hereafter
will assume it to be zero, namely, our calculation will be carried
out in a spatial flat FRW universe \cite{CMB..Komatsu}. And the
function $a(t)$ is the scale factor and its current value is always
set to be unity, therefore the time recording of the history of the
universe can be represented by the redshift $z$ with the relation
$a=(1+z)^{-1}$.

As mentioned in the preceding section, the time derivatives of $a$
are defined as
\begin{equation}
  H(z)\equiv\frac{\dot{a}}{a},
\end{equation}
\begin{equation}\label{Eq:q}
  q(z)\equiv-\frac{1}{H^{2}}\frac{\ddot{a}}{a}=\frac{1}{2}(1+z)\frac{[H(z)^{2}]'}{H(z)^{2}}-1,
\end{equation}
\begin{eqnarray}\label{Eq:jerk}
  j(z)&&\equiv-\frac{1}{H^{3}}\frac{\dddot{a}}{a}= \nonumber \\
  &&-\left[\frac{1}{2}(1+z)^{2}\frac{[H(z)^{2}]''}{H(z)^{2}}-(1+z)\frac{[H(z)^{2}]'}{H(z)^{2}}+1\right],
\end{eqnarray}
where the prime denotes the derivative with respect to the redshift
$z$.

Within the assumption of a constant $j(z)$, Eq.(\ref{Eq:jerk}) is an
Euler equation solved as
\begin{equation}
  H^{2}(z)=\tilde{C_{1}}(1+z)^{\alpha_{+}}+\tilde{C_{2}}(1+z)^{\alpha_{-}},
\end{equation}
where $\tilde{C_{1}}$ and $\tilde{C_{2}}$ are arbitrary constants
and
\begin{equation}
  \alpha_{\pm}=\frac{3\pm\sqrt{1-8j}}{2}.
\end{equation}
The solution gives a constraint that $j<0.125$. With the
redefinition of $C_{1,2}=\tilde{C}_{1,2}/H_{0}^{2}$, where the
subscript '0' stands for the current value of a quantity, the
expansion factor of the FRW cosmology can be written as
\begin{equation}\label{Eq:E}
  E(z)=\frac{H(z)}{H_{0}}=(C_{1}(1+z)^{\alpha_{+}}+C_{2}(1+z)^{\alpha_{-}})^{\frac{1}{2}}.
\end{equation}
Then using the definition of the expansion factor $E(z)=H(z)/H_{0}$,
we can substitute $E(z)$ into Eq.(\ref{Eq:jerk}) and replace $H(z)$
without any essential change.

In order to determine the constants appeared in Eq.(\ref{Eq:E}) and
their physical meanings in the solution, we can apply a particular
cosmological model as a reference. In the $\Lambda$CDM model, we
have $j=-1$ and Eq.(\ref{Eq:E}) becomes
\begin{equation}
  E(z)=(C_{1}(1+z)^3+C_{2})^{\frac{1}{2}}.
\end{equation}
It is clear that $C_{1}$ becomes the matter term $\Omega_{m0}$
(including the ordinary matter and the dark matter) and $C_{2}$
represents the cosmological constant term $\Omega_{\Lambda}$.
However, we should notice that this correspondence is just valid in
the frame of the $\Lambda$CDM model. We can not say for sure the
explicit relationship between $C_{1}$, $C_{2}$ and $\Omega_{m0}$,
$\Omega_{\Lambda}$, but the approximation of $\Lambda$CDM model is a
proper reference for us to find the real meaning of the parameters
$C_{1}$ and $C_{2}$. Additionally, the current value of
Eq.(\ref{Eq:E}) gives
\begin{equation}
  C_{1}+C_{2}=1.
\end{equation}
%In our following calculation, we will apply the above functions in our model.

The assumption of $j=constant$ leads Eq.(\ref{Eq:jerk}) to a
homogeneous Euler equation which can be seen more obviously under
the variable substitutions:
\begin{equation}
  (1+z)\rightarrow x, \qquad  H^{2}\rightarrow y(x).
\end{equation}
This result provides us the possibility to test the deviation of
$j(z)$ from -1 or other value in the past, but the calculation is
dependent on the functional form of $j(z)$. Similarly as the methods
studying the properties of the interaction between dark energy and
dark matter by assuming some phenomenological models
\cite{interaction..Cai}, one possible proposal in reconstructing
jerk is
\begin{equation} \label{Eq:re_j}
  j(z)=j_{0}+j_{1}\frac{f(z)}{E^{2}(z)},
\end{equation}
where $j_{0}$ and $j_{1}$ are constants needed to be constrained
while $f(z)$ is an arbitrary function of redshift $z$, the different
choices of which can lead to different reconstructions of
$j(z)$.This kind of assumption has several advantages. Firstly, it
can make Eq.(\ref{Eq:jerk}) analytically solvable under particular
models of $f(z)$. Secondly, it can also make Eq.(\ref{Eq:jerk}) more
symmetric by satisfying both sides of the equation comprised by a
constant term plus a $E^{-2}$ term.

The simplest form of $f(z)$ is perhaps the linear form which can be
written as $f(z)=z$, which is the first order of the linear
expansion of $f(z)$. Moreover, inspired by reconstructing the
equation of state (EoS) of the dark energy, we propose another model
similarly as the CPL parameterizations
\cite{CPL..Chevallier,CPL..Linder}
\begin{equation}
  f(z)=(1-a)=\frac{z}{1+z}.
\end{equation}
To be more general, we can also apply some functions which are
different from the above. We choose the similar JBP
\cite{JBP..Jassal} (in reconstructing the EoS of dark energy) and
the logarithmic model. To sum up, we apply four models of $f(z)$ in
our calculation. Moreover, as mentioned in the previous section, we
regard the standard $\Lambda$CDM model as the fiducial model and
reconstruct $j(z)$ aiming at measuring the departure of $j$ from -1.
Therefore $j_{0}$ in Eq.(\ref{Eq:re_j}) can be set to -1 and the
second term can be seen as the perturbation. This thought is similar
with the previous work \cite{j..Rapetti}, but the specific method is
different. The authors there adopted the Chebyshev polynomials in
reconstructing $j(z)$ and presented a detailed analysis. In our
calculation, we just parameterize the $j(z)$ phenomenologically and
make the Euler equation solvable. Once the above reconstructing
methods are introduced, we can summarize the four parameterizations
as
\begin{eqnarray}
\text{Model I}   && j(z)=-1+j_{1}\frac{z}{E^{2}(z)} \label{Eq:j1} \\
\text{Model II}  && j(z)=-1+j_{1}\frac{z}{1+z}\frac{1}{E^{2}(z)} \label{Eq:j2} \\
\text{Model III} && j(z)=-1+j_{1}\frac{z}{(1+z)^{2}}\frac{1}{E^{2}(z)}  \label{Eq:j3} \\
\text{Model IV}  && j(z)=-1+j_{1}\frac{\ln(1+z)}{E^{2}(z)}.
\label{Eq:j4}
\end{eqnarray}
One point worth noticing is that all these models have $j(z=0)=-1$.
More discussions of this issue will be given in the following
section. Substituting these equations into Eq.(\ref{Eq:jerk}), we
can obtain the solutions of $E(z)$
\begin{widetext}
\begin{eqnarray}
\text{Model I}    && E^{2}(z)=\frac{1}{3}C_{1}(1+z)^{3}+C_{2}+j_{1}(1+z)-\frac{2}{3}j_{1}\ln(1+z) \label{Eq:E1} \\
\text{Model II}   && E^{2}(z)=\frac{1}{3}C_{1}(1+z)^{3}+C_{2}+\frac{j_{1}}{2(1+z)}+\frac{2}{3}j_{1}\ln(1+z)  \label{Eq:E2} \\
\text{Model III}  && E^{2}(z)=\frac{1}{3}C_{1}(1+z)^{3}+C_{2}+\frac{j_{1}}{5(1+z)^{2}}-\frac{j_{1}}{2(1+z)} \label{Eq:E3} \\
\text{Model IV}   &&
E^{2}(z)=\frac{1}{3}C_{1}(1+z)^{3}+C_{2}+\frac{2}{9}j_{1}\ln(1+z)+\frac{1}{3}j_{1}\ln^{2}(1+z).
\label{Eq:E4}
\end{eqnarray}
\end{widetext}
The coefficients $C_{1}$ and $C_{2}$ arise from the process of
solving Eq.(\ref{Eq:jerk}) which is a second order differential
equation. Another constraint that $E(z=0)=1$ gives a relationship
between the constants $C_{1}$, $C_{2}$ and $j_{1}$
\begin{eqnarray}
\text{Model I}    && C_{2}=1-j_{1}-\frac{1}{3}C_{1}  \\
\text{Model II}   && C_{2}=1-\frac{j_{1}}{2}-\frac{1}{3}C_{1}  \\
\text{Model III}  && C_{2}=1+\frac{3}{10}j_{1}-\frac{1}{3}C_{1}  \\
\text{Model IV}   && C_{2}=1-\frac{1}{3}C_{1}
\end{eqnarray}
Thus each model above has two free parameters ($C_{1},j_{1}$) needed
to be constrained by the observational data.

\subsection{Observational data}

The first observational data sample used in our calculation is the
measurements of Type Ia Supernovae (SNe Ia). This kind of
observation plays an important role in discovering the accelerated
expansion of the universe. Its application in constraining the
cosmological models comes from the distance modulus which is defined
as
\begin{equation}
  \mu(z)=5\log(d_{L}/Mpc)+25,
\end{equation}
where $d_{L}$ is the luminosity distance. In a spatially flat FRW
universe, the luminosity distance of a cosmological source at
redshift $z$ reads as
\begin{equation}
  d_{L}=(1+z)\int_{0}^{z}\frac{dz'}{H(z')}.
\end{equation}
The parameters introduced in the model can be obtained through the
$\chi^{2}$ statistics. In our calculations, we choose the
marginalized nuisance parameter \cite{SNe..Nesseris} for $\chi^{2}$
\begin{equation}
  \chi^{2}_{\text{SNe}}=A-\frac{B^{2}}{C}
\end{equation}
where
\begin{equation}
  A=\sum_{i}\frac{[\mu_{obs}(z_{i})-\mu_{th}(z_{i})]^{2}}{\sigma_{i}^{2}},
\end{equation}
\begin{equation}
  B=\sum_{i}\frac{[\mu_{obs}(z_{i})-\mu_{th}(z_{i})]}{\sigma_{i}^{2}},
\end{equation}
\begin{equation}
  C=\sum_{i}\frac{1}{\sigma_{i}^{2}},
\end{equation}
where $\sigma_{i}$ denotes the $1\sigma$ uncertainty of the $i$th
measurement. The subscripts ``obs'' and ``th'' stand for the
observational and theoretical values of a variable respectively. In
our work, we choose the latest Union2.1 compilation of the SNe data
sample \cite{Union2.1..Suzuki} which contains 580 Type Ia supernovae
observations in the redshift range $0<z<1.414$. On the other hand,
the systematic errors in measuring the luminosity distance should also be
considered. Thus we calculate the constraints of Union2.1 with systematic errors
of the jerk parameterizations as well. The method we used here is the same as suggested in Ref.\cite{OHD..Farooq}.

We also adopt the observational Hubble parameter data (OHD) in our
constraints. It is known that the SNe is powerful in constraining
the cosmological models. However, the integration in its formula
makes it hard to reflect the precise evolution of $H(z)$. Therefore,
the fine structure of the expansion history of the universe can be
well indicated by the $H(z)$ data. The measurement of OHD can be
derived from the
differential of redshift $z$ with respect to the cosmic time $t$ %\cite{OHD..Jimenez}
\begin{equation}
  H(z)=-\frac{1}{1+z}\frac{dz}{dt}.
\end{equation}
In this work, we use the 21 $H(z)$ measurements of Ref.
\cite{OHD..Jimenez,OHD..Jimenez2,OHD..Simon,OHD..Stern2,OHD..Gaztanaga,OHD..Moresco}
to constrain the jerk models.

The $\chi^{2}$ value for the OHD can be expressed as
\begin{equation}
  \chi^{2}_{\text{OHD}}=\sum_{i=1}^{21}\frac{[H_{obs}(z_{i})-H_{th}(z_{i})]^{2}}{\sigma_{i}^{2}}.
\end{equation}
The best-fit values of the parameters can be obtained by minimizing
the above quantity. It should be noticed that there is a nuisance
parameter in the OHD constraint: $H_{0}$. Therefore, we marginalize
it by using a prior $H_{0}=74.3\pm2.1$ km s$^{-1}$ Mpc$^{-1}$
\cite{OHD..Freedman} which contains a 2.8\% systematic uncertainty.
On the other hand, the prior value of $H_{0}=68\pm2.8$km s$^{-1}$ Mpc$^{-1}$
was also used in the previous works and showed efficient constraints on the
cosmological models\cite{OHD..Chen}. Except that, this value is more consistent with the newly
released Planck results\cite{Planck}. So we also use this prior in our calculation
and the results may provide valuable comparisons.

Additionally, as the newly measurement at $z=2.3$ from the Baryon
Acoustic Oscillation (BAO) in the Ly$\alpha$ is discovered
\cite{OHD..Busca}, we also adopt this data in our calculation in order
to find the effect of this addition in constraining cosmological models.
For convenience, we will use the term "$H_{2.3}$" in the following
sections to denote this measurement.

\section{Constraint results}\label{constraint}

%===================figure1=======================
\begin{figure*}
 \center
\includegraphics[scale=0.6]{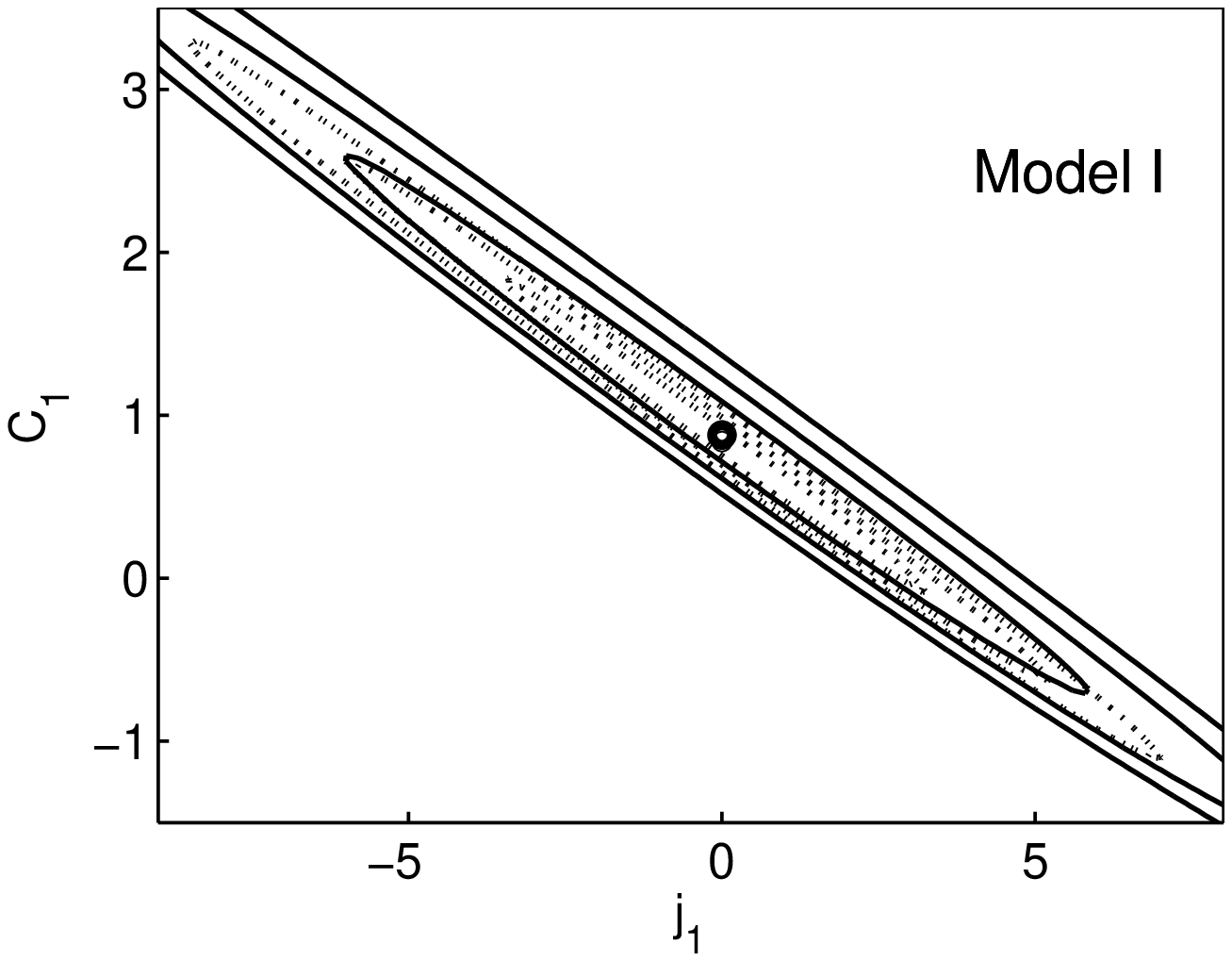}
\includegraphics[scale=0.6]{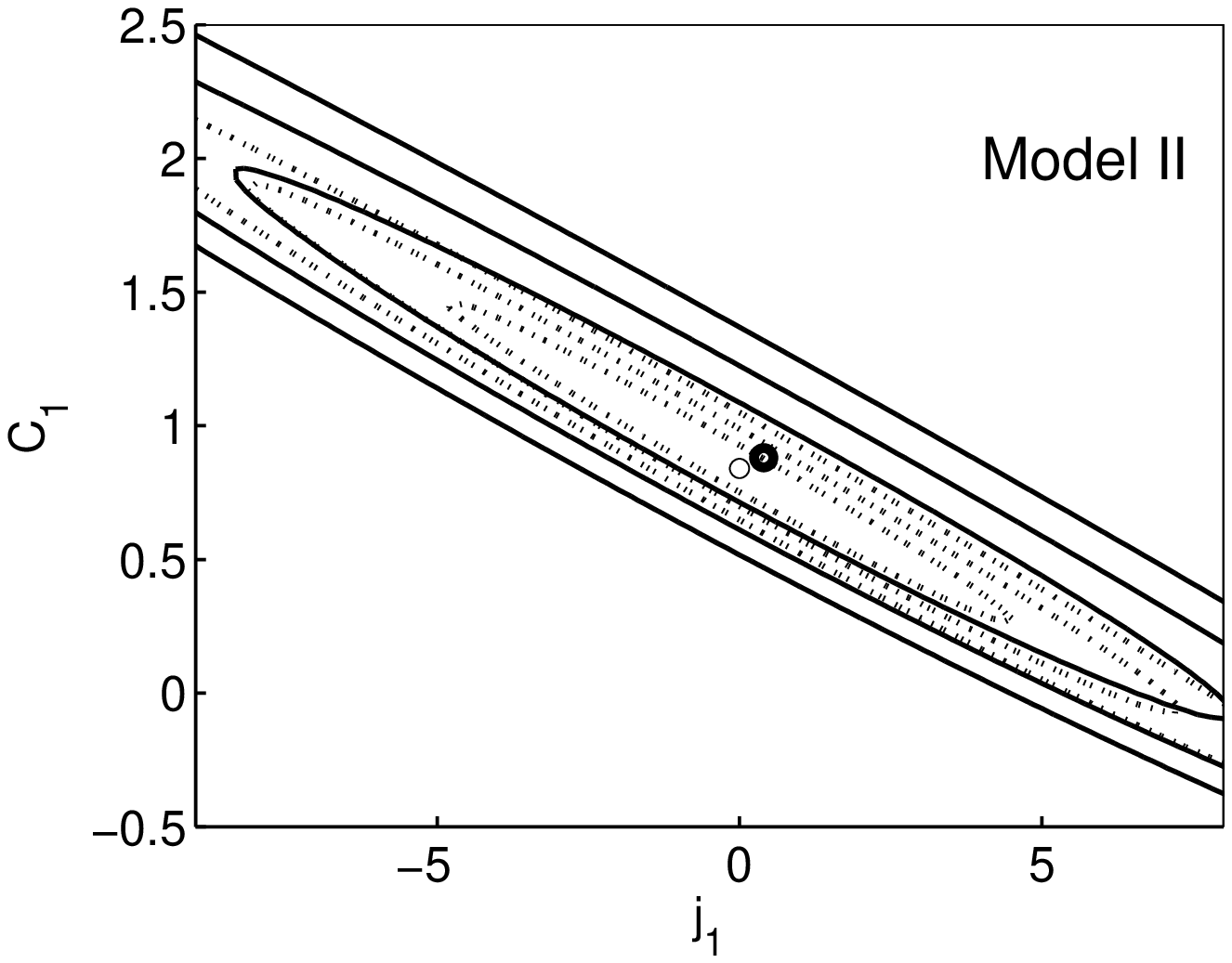}
\includegraphics[scale=0.6]{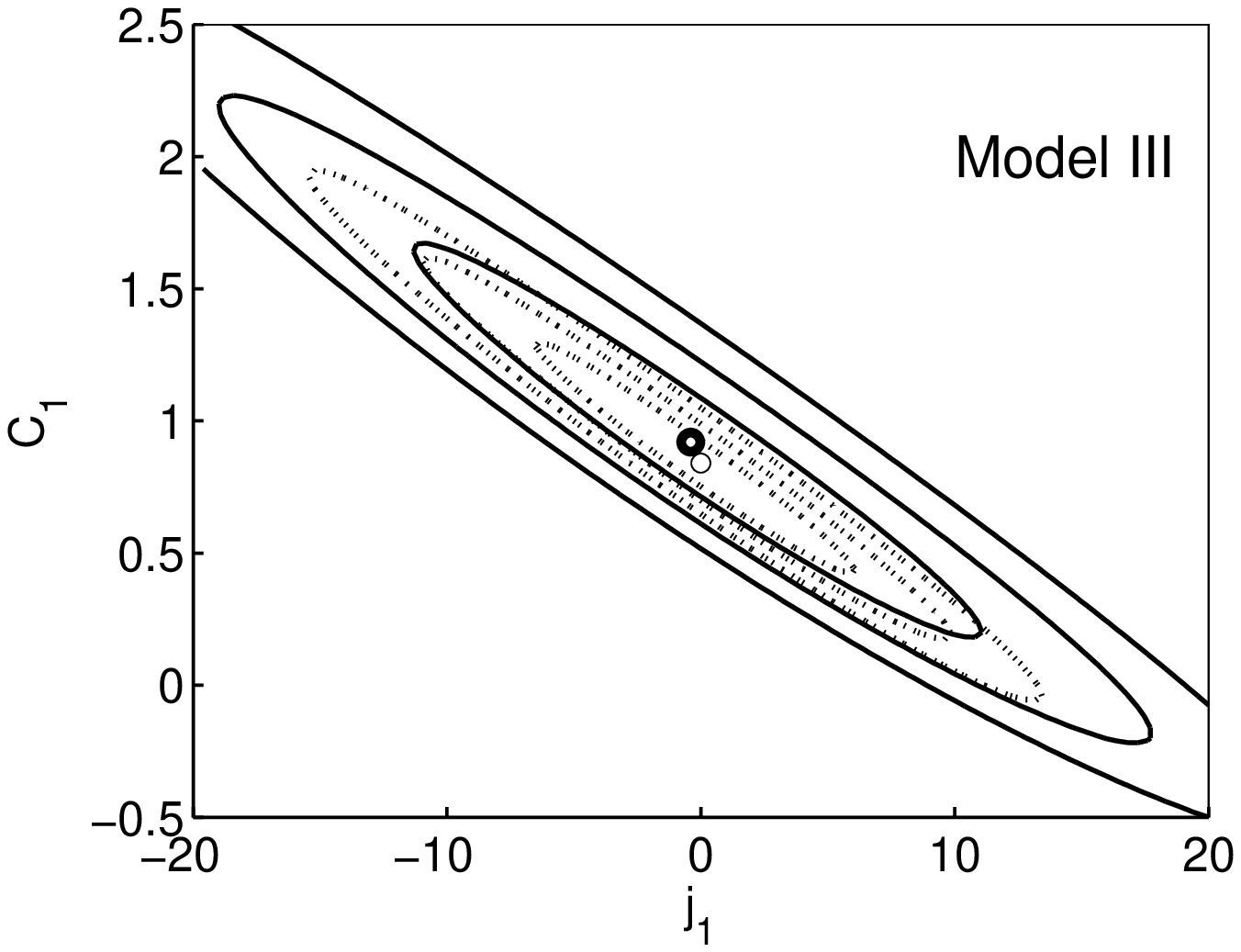}
\includegraphics[scale=0.6]{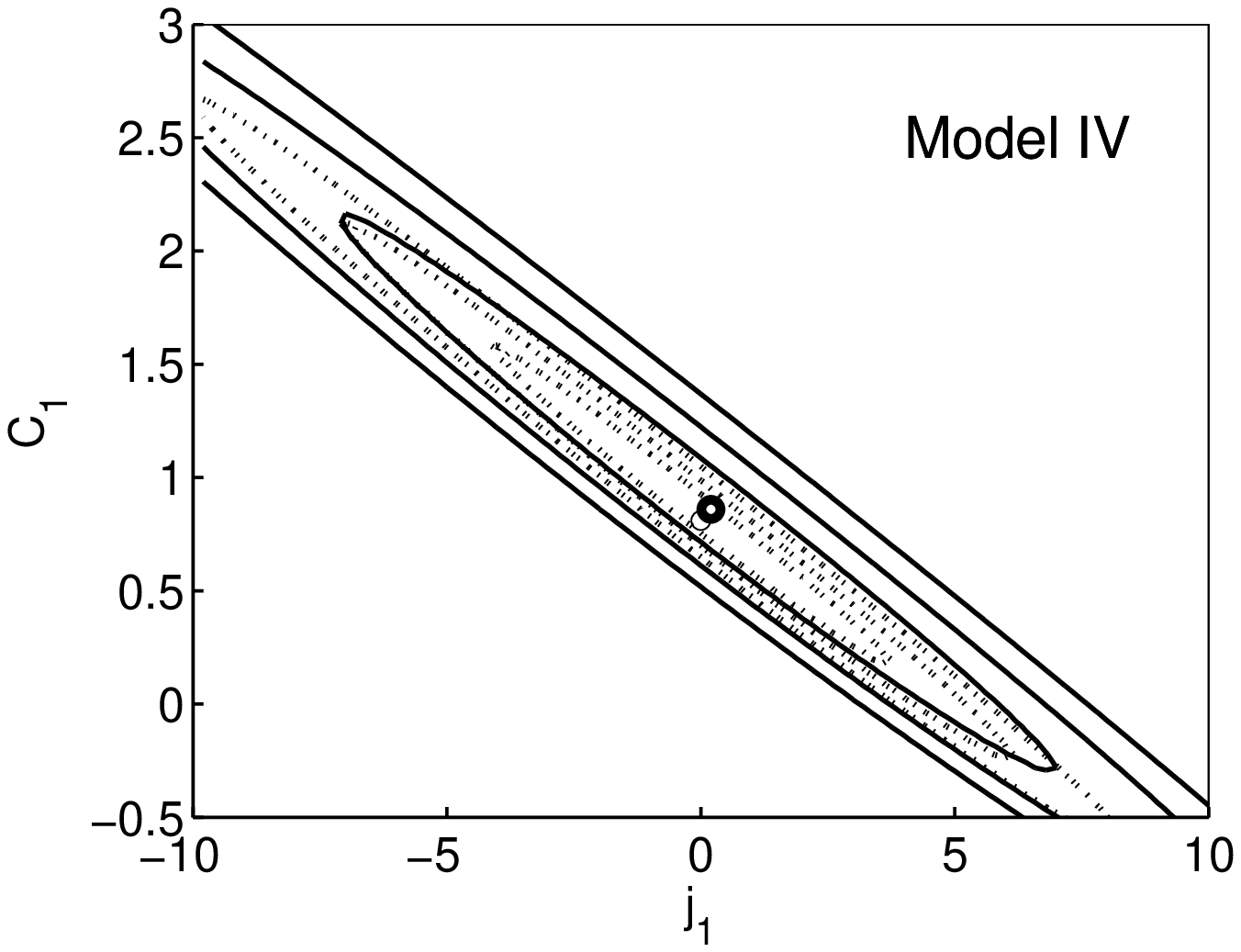}
\caption{The confidence regions of $(j_{1},C_{1})$ obtained from SNe with (solid)
and without (dotted) systematic errors respectively. The 68.3\%, 95.4\% and 99.7\%
confidence level are presented from inner to outer.
The thin (without systematic errors) and thick circles
(with systematic errors) stand for the best-fit values.} \label{fig:cons_SNe}
\end{figure*}
%===================================================
\begin{table*}
\begin{center}\begin{tabular}{l||c|c|c|c|c}
 %\MC{3}{c}{\text{caption}}\\[5pt]
 \hline
               &      data        &        $j_{1}$        &         $C_{1}$         &      $\chi_{\text{min}}^{2}$       &      FoM                       \\ \hline
 %\MC{3}{|c|c|}{\ZZ{-8pt}{15pt}\hfill\normalsize   \hfill  \hfill\normalsize MGCDM     \hfill\normalsize $\Lambda$CDM  }\\ \hline
 %\ZZ{-6pt}{22pt}
 \normalsize Model I          &  SNe $_{a}$               &  $0\pm2.14$          &   $0.84\pm0.60$   &   $561.73$    &  7.7829  \\ \hline
 \normalsize Model I          &  SNe $_{b}$                  &  $0\pm3.68$          &   $0.88\pm1.05$   &   $545.84$    &  2.0923  \\ \hline
 \normalsize Model II         &  SNe $_{a}$               &  $0\pm2.96$          &   $0.84\pm0.37$   &   $561.73$    &  5.5381  \\ \hline
 \normalsize Model II         &  SNe $_{b}$                  &  $0.4\pm5.14$        &   $0.84\pm0.62$   &   $545.83$    &  1.4662  \\ \hline
 \normalsize Model III        &  SNe $_{a}$               &  $0\pm4.03$          &   $0.84\pm0.27$   &   $561.73$    &  4.1031  \\ \hline
 \normalsize Model III        &  SNe $_{b}$                  &  $-0.4\pm7.56$       &   $0.92\pm0.49$   &   $545.84$    &  1.0994  \\ \hline
 \normalsize Model IV         &  SNe $_{a}$               &  $0\pm2.45$          &   $0.81\pm0.41$   &   $561.73$    &  6.5553  \\ \hline
 \normalsize Model IV         &  SNe $_{b}$                  &  $0.2\pm4.49$        &   $0.86\pm0.76$   &   $545.83$    &  1.7271  \\ \hline
\end{tabular}
\end{center}
\caption{The constraint results of the parameters from SNe sample, including the
best-fit values with $1\sigma$ errors of the parameters and the FoM
of four jerk parameterizations. (the subscript $a$: without systematic errors; $b$: with systematic errors.)}\label{tab:SNe}
\end{table*}
%===================figure2=======================
\begin{figure*}
 \center
\includegraphics[scale=0.6]{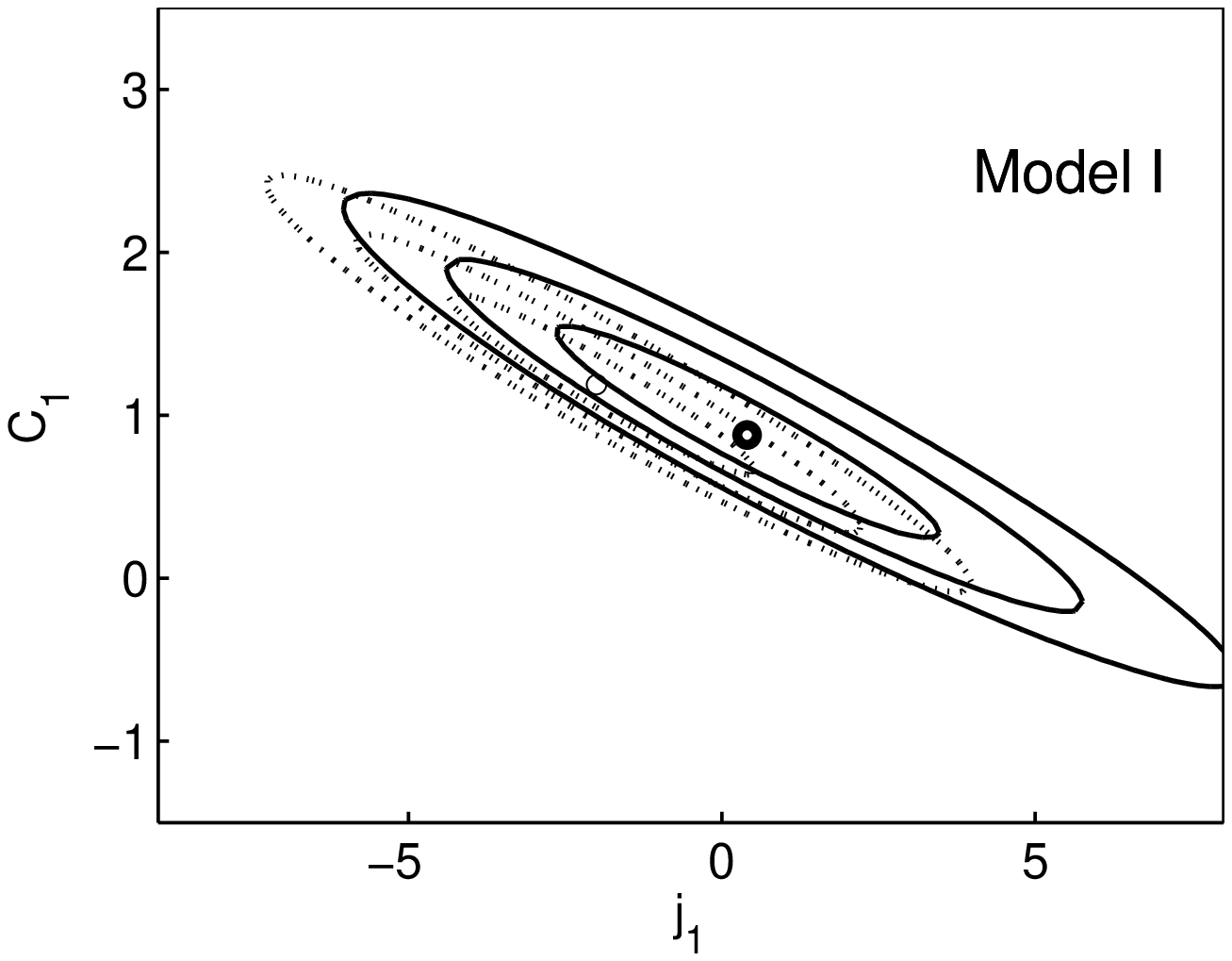}
\includegraphics[scale=0.6]{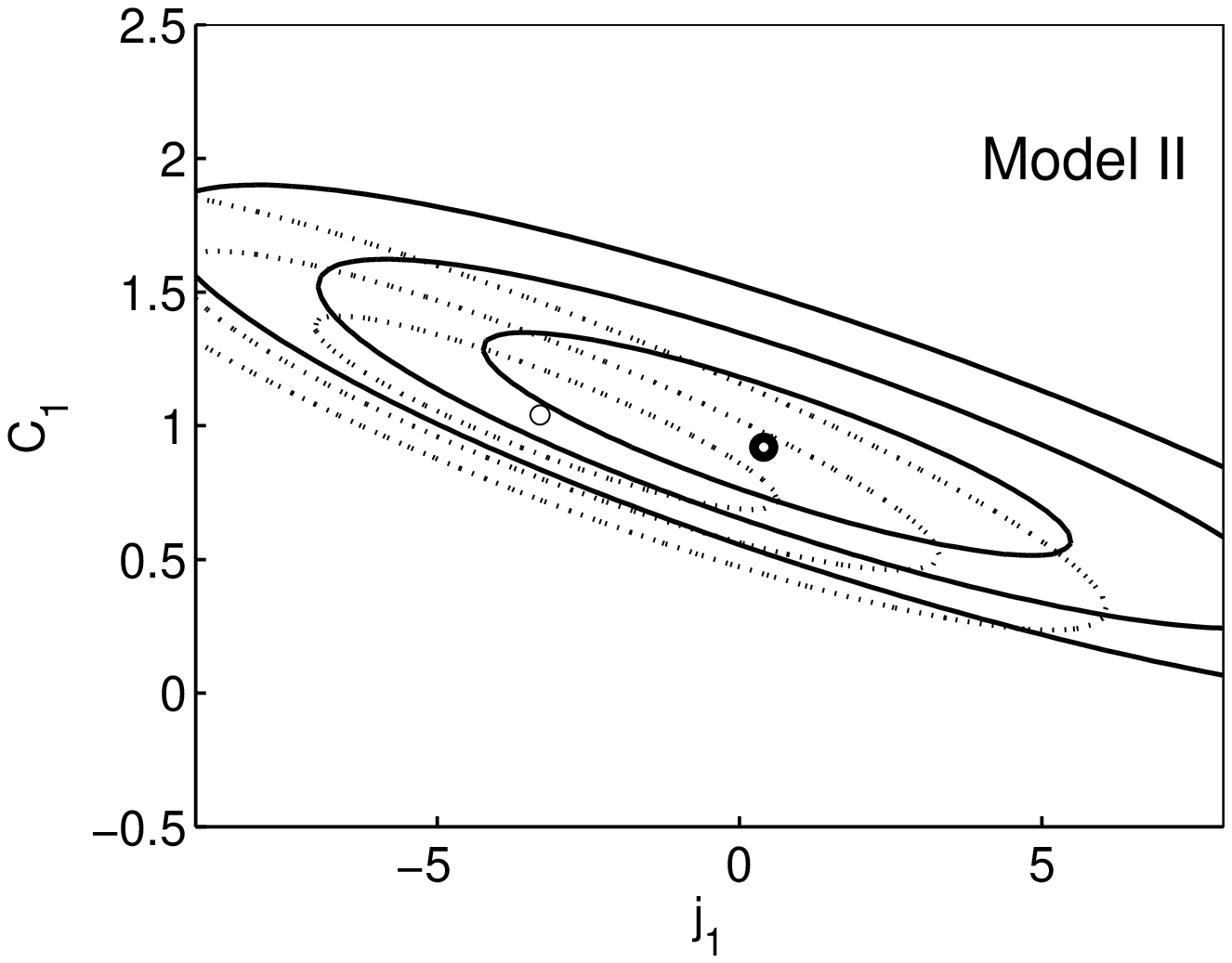}
\includegraphics[scale=0.6]{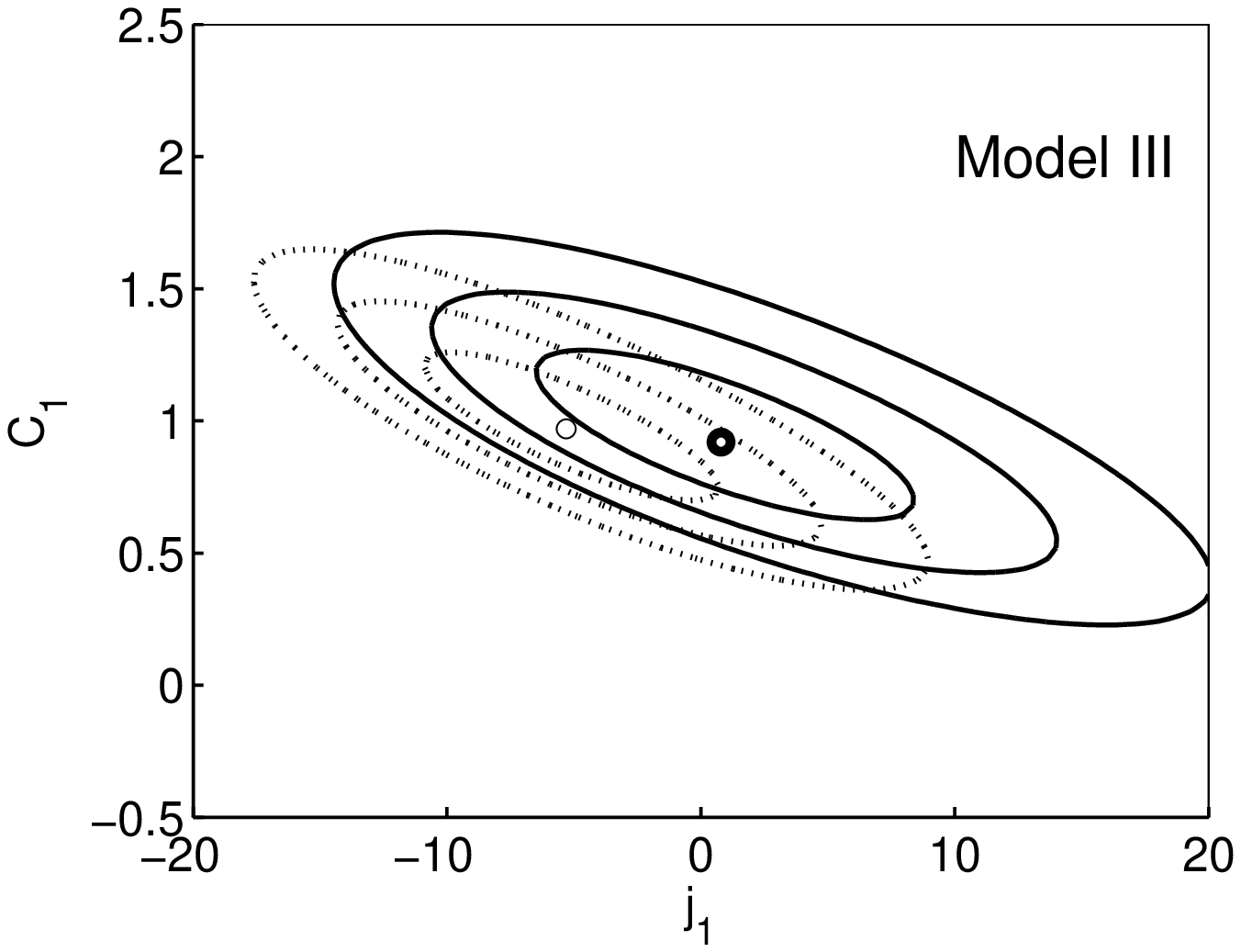}
\includegraphics[scale=0.6]{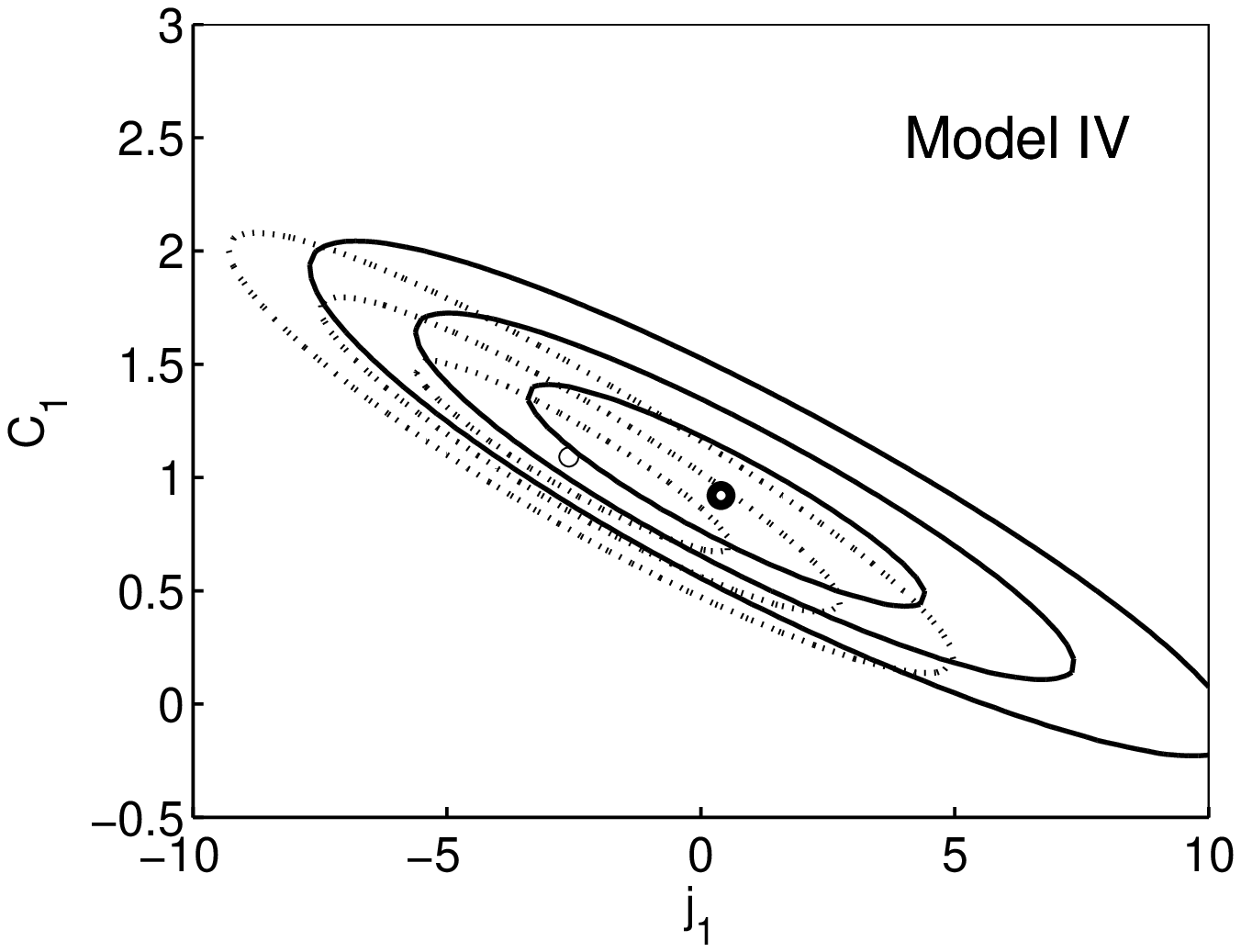}
\caption{The confidence regions of $(j_{1},C_{1})$ obtained from OHD with $H_{0}$ priors of
$H_{0}=74.3\pm2.1$ km s$^{-1}$ Mpc$^{-1}$ (dotted) and $H_{0}=68\pm2.8$ km s$^{-1}$ Mpc$^{-1}$ (solid)
respectively. The 68.3\%, 95.4\% and 99.7\%
confidence level are presented from inner to outer.
The thin ($H_{0}=74.3\pm2.1$ km s$^{-1}$ Mpc$^{-1}$) and thick circles
($H_{0}=68\pm2.8$ km s$^{-1}$ Mpc$^{-1}$) stand for the best-fit values.} \label{fig:cons_OHD}
\end{figure*}
%=================================================
\begin{table*}
\begin{center}\begin{tabular}{l||c|c|c|c|c}
 %\MC{3}{c}{\text{caption}}\\[5pt]
 \hline
               &      data        &        $j_{1}$        &         $C_{1}$         &      $\chi_{\text{min}}^{2}$       &      FoM                       \\ \hline
 %\MC{3}{|c|c|}{\ZZ{-8pt}{15pt}\hfill\normalsize   \hfill  \hfill\normalsize MGCDM     \hfill\normalsize $\Lambda$CDM  }\\ \hline
 %\ZZ{-6pt}{22pt}
 \normalsize Model I          &  OHD (H$_{0}'$)           &  $-2.0\pm1.63$   &  $1.19\pm0.36$   &   $20.65 $    &  5.8160  \\ \hline
 \normalsize Model I          &  OHD (H$_{0}$)            &  $0.4\pm2.02$    &  $0.88\pm0.42$   &   $19.85 $    &  3.5082  \\ \hline
 \normalsize Model II         &  OHD (H$_{0}'$)           &  $-3.3\pm2.60$   &  $1.04\pm0.24$   &   $20.44 $    &  3.7051  \\ \hline
 \normalsize Model II         &  OHD (H$_{0}$)            &  $0.4\pm3.35$    &  $0.92\pm0.27$   &   $19.85 $    &  2.2067  \\ \hline
 \normalsize Model III        &  OHD (H$_{0}'$)           &  $-5.3\pm3.92$   &  $0.97\pm0.18$   &   $20.27 $    &  2.4601  \\ \hline
 \normalsize Model III        &  OHD (H$_{0}$)            &  $0.8\pm4.99$    &  $0.92\pm0.19$   &   $19.86 $    &  1.4410  \\ \hline
 \normalsize Model IV         &  OHD (H$_{0}'$)           &  $-2.6\pm2.06$   &  $1.09\pm0.28$   &   $20.53 $    &  4.5895  \\ \hline
 \normalsize Model IV         &  OHD (H$_{0}$)            &  $0.4\pm2.64$    &  $0.92\pm0.32$   &   $19.85 $    &  2.7454  \\ \hline
\end{tabular}
\end{center}
\caption{The constraint results of the parameters from OHD sample, including the
best-fit values with $1\sigma$ errors of the parameters and the FoM
of four jerk parameterizations. ($H_{0}'$ denotes the value $74.3\pm2.1$ km s$^{-1}$ Mpc$^{-1}$ of the prior is used, while $H_{0}$
denotes $68\pm2.8$ km s$^{-1}$ Mpc$^{-1}$ is applied.) }\label{tab:OHD}
\end{table*}

%===================figure3=======================
\begin{figure*}
 \center
\includegraphics[scale=0.6]{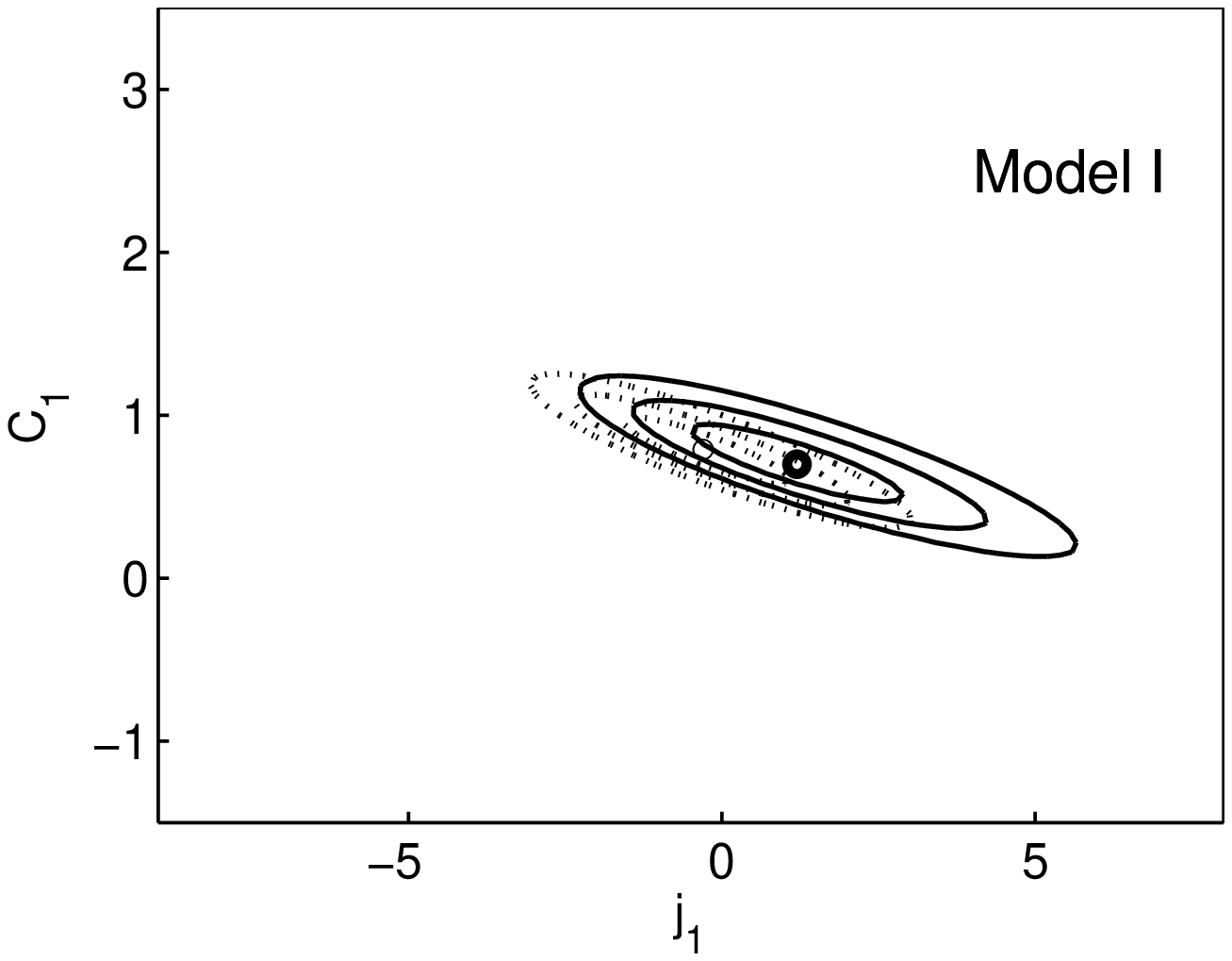}
\includegraphics[scale=0.6]{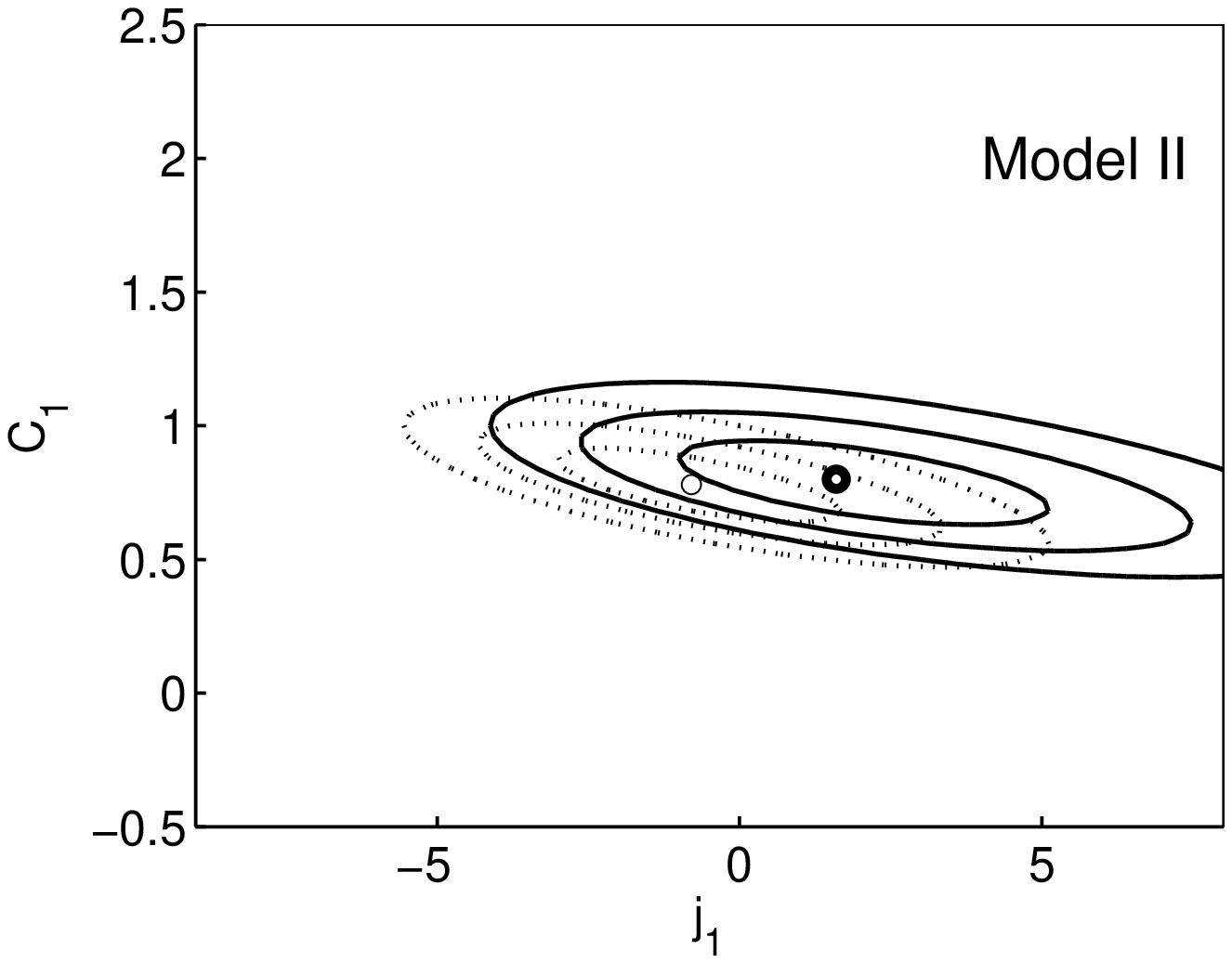}
\includegraphics[scale=0.6]{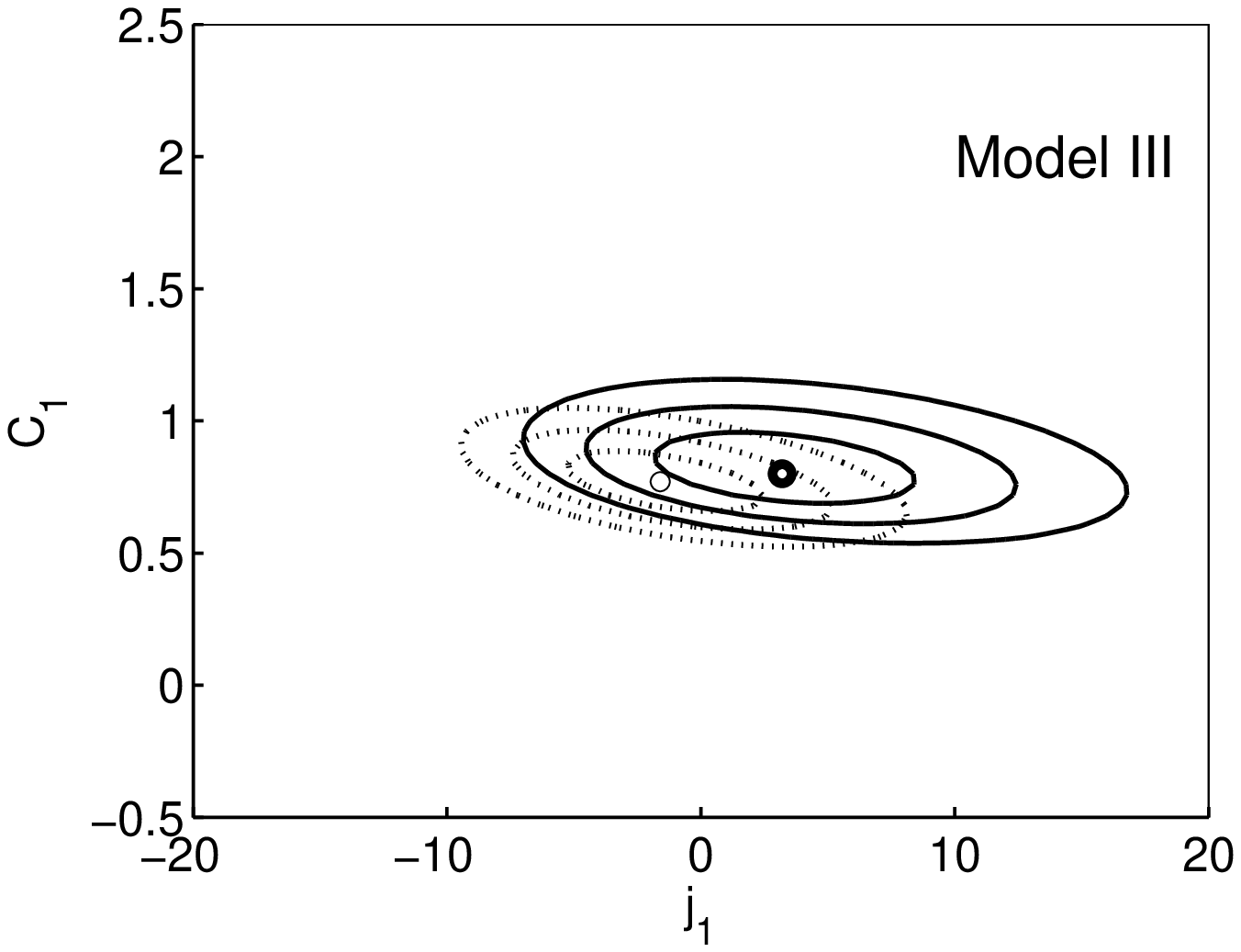}
\includegraphics[scale=0.6]{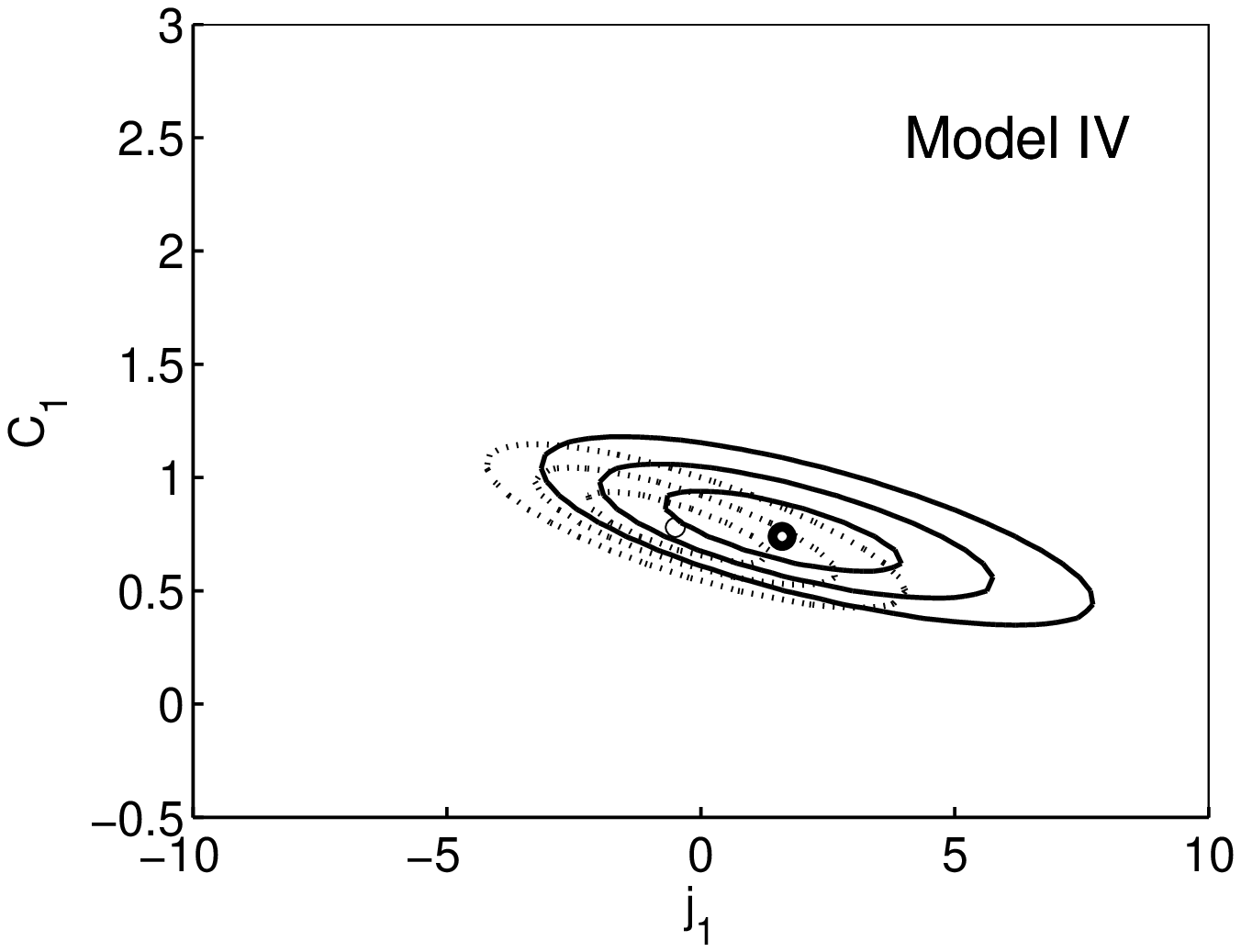}
\caption{The confidence regions of $(j_{1},C_{1})$ obtained from OHD+H$_{2.3}$ with $H_{0}$ priors of
$H_{0}=74.3\pm2.1$ km s$^{-1}$ Mpc$^{-1}$ (dotted) and $H_{0}=68\pm2.8$ km s$^{-1}$ Mpc$^{-1}$ (solid)
respectively. The 68.3\%, 95.4\% and 99.7\%
confidence level are presented from inner to outer.
The thin ($H_{0}=74.3\pm2.1$ km s$^{-1}$ Mpc$^{-1}$) and thick circles
($H_{0}=68\pm2.8$ km s$^{-1}$ Mpc$^{-1}$) stand for the best-fit values.} \label{fig:cons_OHD_N}
\end{figure*}
%==================================================
\begin{table*}
\begin{center}\begin{tabular}{l||c|c|c|c|c}
 %\MC{3}{c}{\text{caption}}\\[5pt]
 \hline
               &      data        &        $j_{1}$        &         $C_{1}$         &      $\chi_{\text{min}}^{2}$       &      FoM                       \\ \hline
 %\MC{3}{|c|c|}{\ZZ{-8pt}{15pt}\hfill\normalsize   \hfill  \hfill\normalsize MGCDM     \hfill\normalsize $\Lambda$CDM  }\\ \hline
 %\ZZ{-6pt}{22pt}
 \normalsize Model I          &  OHD+$H_{2.3}$ (H$_{0}'$)  &  $-0.3\pm0.87$   &  $0.79\pm0.22$   &   $22.24 $    &  17.488  \\ \hline
 \normalsize Model I          &  OHD+$H_{2.3}$ (H$_{0}$)   &  $1.2\pm1.11$    &  $0.70\pm0.15$   &   $20.30 $    &  10.5487  \\ \hline
 \normalsize Model II         &  OHD+$H_{2.3}$ (H$_{0}'$)  &  $-0.8\pm1.60$   &  $0.78\pm0.09$   &   $22.12 $    &  9.8872  \\ \hline
 \normalsize Model II         &  OHD+$H_{2.3}$ (H$_{0}$)   &  $1.6\pm2.31$    &  $0.80\pm0.11$   &   $20.41 $    &  5.8390  \\ \hline
 \normalsize Model III        &  OHD+$H_{2.3}$ (H$_{0}'$)  &  $-1.6\pm2.64$   &  $0.77\pm0.07$   &   $21.98 $    &  6.0180  \\ \hline
 \normalsize Model III        &  OHD+$H_{2.3}$ (H$_{0}$)   &  $3.2\pm3.35$    &  $0.80\pm0.07$   &   $20.52 $    &  3.4702  \\ \hline
 \normalsize Model IV         &  OHD+$H_{2.3}$ (H$_{0}'$)  &  $-0.5\pm1.21$   &  $0.78\pm0.10$   &   $22.18 $    &  12.9475 \\ \hline
 \normalsize Model IV         &  OHD+$H_{2.3}$ (H$_{0}$)   &  $1.6\pm1.55$    &  $0.74\pm0.10$   &   $20.38 $    &  7.6735 \\ \hline
\end{tabular}
\end{center}
\caption{The constraint results of the parameters from OHD+H$_{2.3}$ sample, including the
best-fit values with $1\sigma$ errors of the parameters and the FoM
of four jerk parameterizations.($H_{0}'$ denotes the value $74.3\pm2.1$ km s$^{-1}$ Mpc$^{-1}$ of the prior is used, while $H_{0}$
denotes $68\pm2.8$ km s$^{-1}$ Mpc$^{-1}$ is applied.) }\label{tab:OHD_N}
\end{table*}

\subsection{Constraints from SNe and OHD}

Our constraint results are presented in Fig.\ref{fig:cons_SNe} and Fig.\ref{fig:cons_OHD}where
the $1\sigma$, $2\sigma$ and $3\sigma$ confidence regions are shown.
These are obtained by finding the contours of $\chi^{2}_{min}+2.3, 6.17, 11.8$
in the parameter space respectively.
Also, we summarize the best-fit values and the corresponding
uncertainties of the parameters in Table.\ref{tab:SNe} and Table.\ref{tab:OHD}. From
these results, we can see that the $\Lambda$CDM model or the
$j(z)=-1$ model is well accommodated by supernovae observations (with and without systematic errors). The
best-fit values of $j_{1}$ in four jerk models are very small which
imply that the perturbation term in Eq.(\ref{Eq:j1})-(\ref{Eq:j4})
can be ignored. Therefore, the standard $\Lambda$CDM model is well
preferred by SNe data. The difference between the samples considering and not considering systematic
errors is also obvious. The constraints from the former ones are apparently looser than the latter ones.
The 2$\sigma$ confidence regions of SNe not considering systematic errors is almost overlap with the 1$\sigma$ confidence
regions of considering systematic errors ones. This situation is expectable since the consideration of the systematic errors
means the reduction of the accuracy of the information we obtained. And this can be well reflected by the confidence regions of the
parameters.

On the other hand, the OHD constraints show different results about $\Lambda$CDM model.
All the four jerk models indicate a tendency of deviation
of the universe from the $\Lambda$CDM model. The best-fit values of
$j_{1}$ given by OHD are less than zero and can not be neglected
when the first prior $H_{0}=74.3\pm2.1$ km s$^{-1}$ Mpc$^{-1}$ is used .
Moreover, these situation appears in all the four models. However,
the choice of the second prior $H_{0}=68\pm2.8$ km s$^{-1}$ Mpc$^{-1}$ changes the above situation.
All the four jerk models show the best-fit values of $j_{1}\geq0$. These results are more
consistent with the SNe ones. But we should notice that the second prior of $H_{0}$ gives worse constraints of the
parameters than the first one.

Based on these results, we can hardly say that the OHD and SNe give very different
constraints because the uncertainties should be taken into account.
The confidence regions of these two data samples show the overlap at
certain confidence level and this phenomenon can be seen as a signal
that the OHD and SNe can give similar constraints. In particular, the best-fit values of the constraint from one data sample can
locate in the 1$\sigma$ confidence region using different sample.
Except that, the tendencies of the confidence regions
given by these two data samples are very similar. Therefore, the
present results approve the previous works that the OHD can play the
same role as SNe in constraining cosmological models
\cite{OHD..Zhai,OHD..Lin}.

An important point worth noticing is the differences of the powers
of OHD and SNe in constraining the cosmological models. In the jerk
parameterizations, the OHD shows that the uncertainties of the
parameters are smaller than the SNe ones. Thus we may conclude that
the OHD is more powerful than SNe. Because of the smaller size of
OHD sample, this is quite a satisfactory phenomenon. However, looking
at the confidence regions, we find that the constraints given by OHD
are not as strict as the SNe ones.

Except for the uncertainties of the parameters, the size of the
confidence region at certain level is another useful tool in
evaluating the powers of observational data in constraining
cosmological models. In order to compare the abilities of OHD and
SNe in constraining the jerk models quantitatively, we adopt the
test of Figure of Merit (FoM). Similarly, we choose the definition
of FoM as the Dark Energy Task Force (DETF) used
\cite{FoM..Albrecht,FoM..Huterer}. The FoM is the inverse of the
area of the 95.4\% confidence level region $A_{95}$ in the parameter
space (the $j_{1}-C_{1}$ plane in our models.). Once the
normalization is considered, we define FoM as
\cite{FoM..Mortonson,FoM..Ruiz}
\begin{equation}
  \text{FoM}_{(j_{1},C_{1})}\approx\frac{6.17\pi}{A_{95}}.
\end{equation}
If the probability distribution of the parameter is Gaussian, the
approximate equality in this equation becomes exact.

The FoM results are summarized in the rightmost column of
Table.\ref{tab:SNe},Table.\ref{tab:OHD} and Table.\ref{tab:OHD_N}. We find that the FoM test does not clearly approve
the above conjecture that OHD is superior than SNe. The FoM of SNe without considering the systematic errors
is about 1.5 times of the OHD ones. But the addition of the systematic errors changes this
because its FoM is smaller than OHD.
Thus in fact, the constraints in all the four models show that the powers of OHD and SNe in constraining cosmological models
are hard to evaluate because their FoM values are sensitive to the choice of systematic errors. About the Hubble parameter data themselves,
their constraints are also sensitive to the choice of the prior. The second prior $H_{0}=68\pm2.8$ km s$^{-1}$ Mpc$^{-1}$ gives
worse constraints as we mentioned in the previous paragraphs.

We can conclude that the SNe data give both larger
uncertainties of the parameters and strict constraint. This seems to
be paradoxical but in fact not, because it reflects that the
correlations of the parameters given by OHD is weaker than SNe
gives. In other words, the parameters in OHD are more independent
between each other \cite{FoM..Coe}.

\begin{figure*}
%\vspace{.2in}
\centerline{\psfig{figure=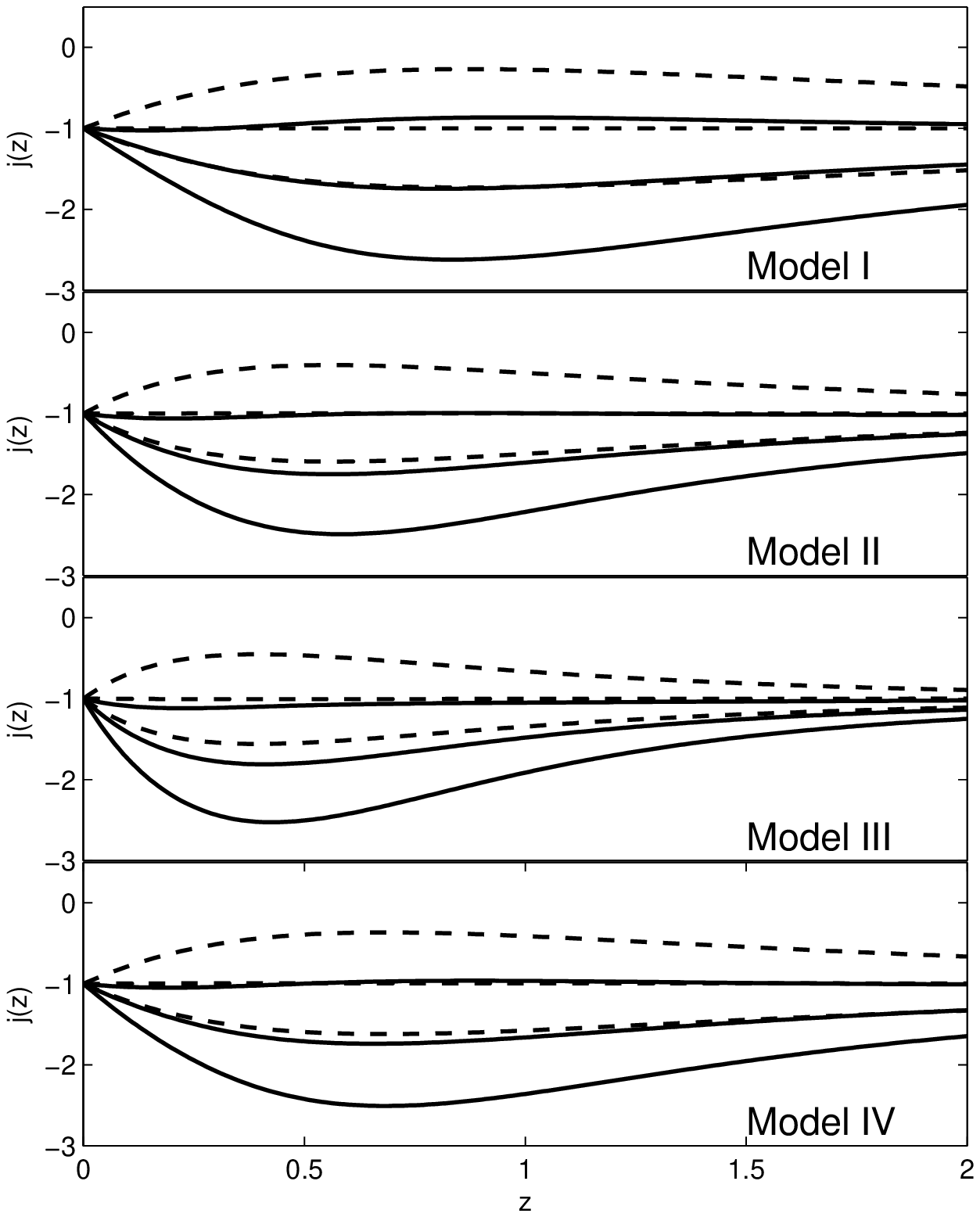,width=3.5truein,height=6.5truein}
\psfig{figure=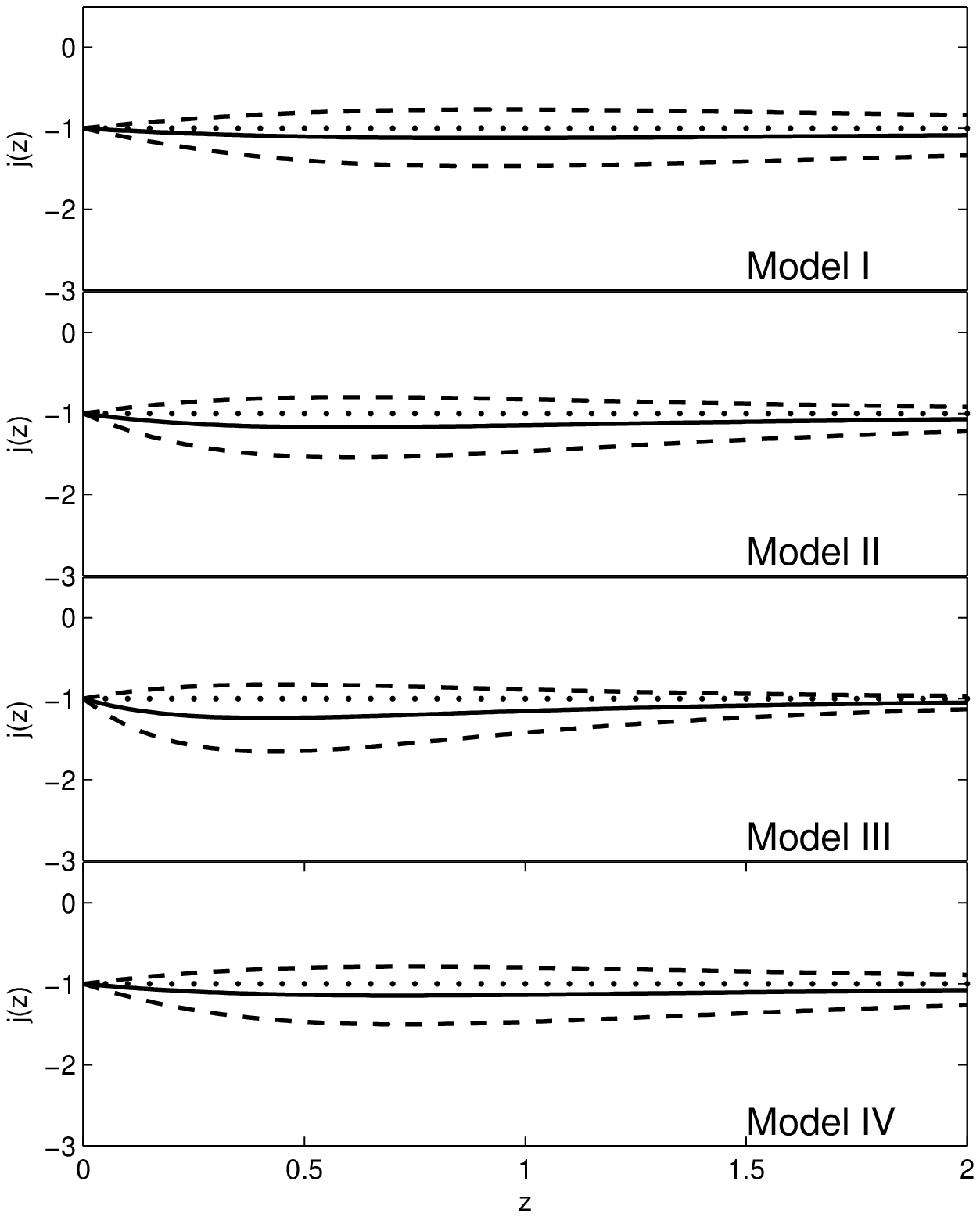,width=3.5truein,height=6.5truein}
\hskip 0.01in} \caption{The reconstruction of jerk parameter. $Left$: The dotted and solid lines represent the best-fit and $1\sigma$ error curves of SNe and OHD
respectively. Right: The results obtained from OHD+H2.3 including the best-fit
value and 1$\sigma$ error. The dots stand for j(z) = -1 line.}\label{fig:recon_j}
\end{figure*}

The reconstructions of $j(z)$ of the jerk models are plotted in the
left panel of Fig.\ref{fig:recon_j} as illustrations (The best-fit values and errors of the parameters are chosen
by SNe (without systematic errors) and OHD (first prior)). The SNe data strongly favor the
$\Lambda$CDM model, while the OHD prefers a deviation. Once the
uncertainties are considered, this deviation disappeared. The curve
of standard $\Lambda$CDM model evolves almost along the boundary of
the $1\sigma$ confidence level.

%===================figure4=======================
%=================================================

\subsection{The addition of $H_{2.3}$}

Recently, the new OHD measurement at $z=2.3$ from BAO in Ly$\alpha$
was discovered with the value of $H(z=2.3)=224\pm8$ km s$^{-1}$
Mpc$^{-1}$ \cite{OHD..Busca}. This measurement has been used in
constraining cosmological models \cite{OHD..Farooq2} and shows that
the addition of it can provide restrictive constraint. In
particular, the constraints are tighter than those from SNe data
\cite{OHD..Farooq2}. This should be attributed to the high redshift
of this measurement and the apparently small uncertainty which has
been carefully estimated. This can naturally increase its weight in
the $\chi^{2}$ statistics.

In our calculation, we also adopt this measurement to the OHD
sample. The constraint results of the jerk parameterizations are
presented in Fig.\ref{fig:cons_OHD_N}. The best-fit values and the
uncertainties of the parameters are also summarized in
Table.\ref{tab:OHD_N}. We can find that similarly, the addition of $H_{2.3}$
improves the constraints apparently. This can be obtained from both
the uncertainties and FoM test, even it can help OHD to provide
tighter constraints than SNe data. The values of FoM based on OHD
are almost double of that from SNe, which is a significant progress
and is consistent with the previous work \cite{OHD..Farooq2}. Also,
the best-fit values of the parameters show a preference to the
$\Lambda$CDM model, where the absolute value of $j_{1}$ reduces.
Taking into account of the uncertainties, we find that OHD+H$_{2.3}$
give more consistent constraints compared with the SNe ones

The reconstruction of $j(z)$ are plotted in the right panel of
Fig.\ref{fig:recon_j} (parameters applied are the same as left panel). It can be found that the standard
$\Lambda$CDM model is apparently favored by OHD+H$_{2.3}$.
Additionally, the $1\sigma$ error of the reconstruction is also
reduced even smaller than the SNe ones. It is anticipated that
OHD+H$_{2.3}$ provides tighter and more efficient constraint than
SNe.

Except that, we can see that all the data samples including SNe, OHD
and OHD+H$_{2.3}$ show that $j(z)$ goes to -1 as the redshift $z$
increase. This is natural as we go back along the cosmic evolution
there was a matter-dominated phase before the current accelerated
expansion. The Hubble parameter can be approximated as
\begin{equation}
  H^{2}(z)\approx\Omega_{m0}(1+z)^{3}.
\end{equation}
Substituting this equation into Eq.(\ref{Eq:jerk}), we may find the
first two terms in the bracket of the right hand side cancel out and
just leave $j(z)=-1$. Perhaps this can be treated as a criteria in
reconstructing $j(z)$ because the matter-dominated phase is almost a
necessity.

\subsection{The Hubble parameter and the equation of state}
%===================figure4=======================
\begin{figure*}
%\vspace{.2in}
\centerline{\psfig{figure=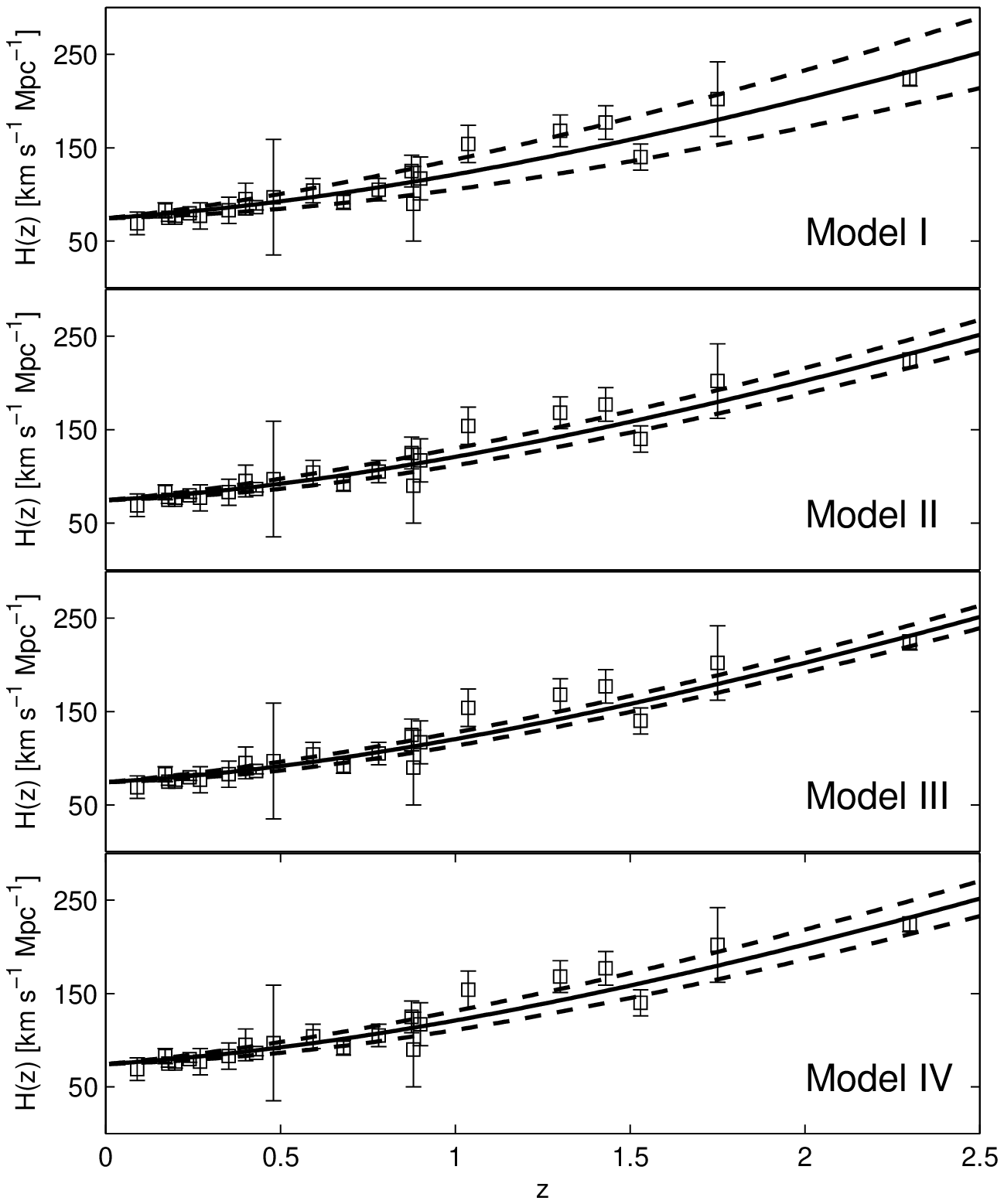,width=3.5truein,height=6.5truein}
\psfig{figure=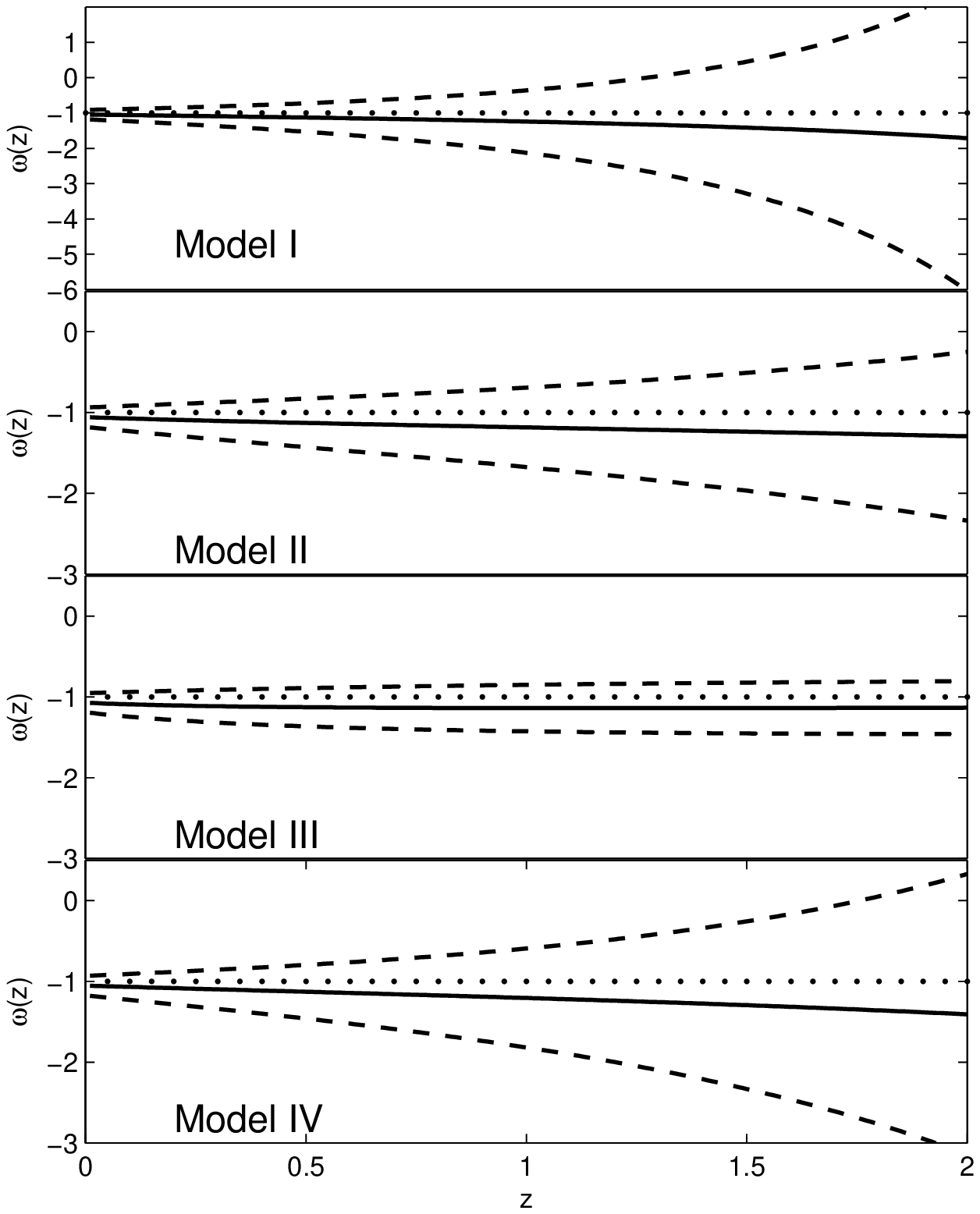,width=3.5truein,height=6.5truein}
\hskip 0.01in} \caption{$Left$: The evolution of the Hubble
parameter $H(z)$ with respect to the redshift $z$. The solid curves
represent the best fit results of OHD+H$_{2.3}$ while the dashed
lines stand for the 1$\sigma$ errors. The original data points are
also shown. $Right$: The reconstruction of the equation of state as
a function of the redshift $z$ by the use of the OHD+H$_{2.3}$
results. The solid lines and dashed lines stand for the best-fit
results and 1$\sigma$ errors respectively. The dotted lines
represent the $\Lambda$CDM case.}\label{fig:recon_H}
\end{figure*}
%=================================================

As we know, the jerk parameter relates to the third time derivative
of the scale factor. Although $j(z)$ is a good function to describe
the evolution of the universe, to identify different dark energy
models as a part of the so-called "statefinder" diagnostic, to study
the future including the type of the singularities of the universe
\cite{future..Dabrowski,future..Dabrowski2,future..Astashenok,future..Frampton,future..Frampton2}
and so forth, its dynamical properties are not easily handled, at
least not as obvious as the deceleration factor or the Hubble
parameter. In order to study the effects of our reconstructions of
$j(z)$ and trace its behavior, we calculate the Hubble parameter and
the equation of state in this section.

In the left pannel of Fig.\ref{fig:recon_H}, we plot the evolution
of the Hubble parameter $H(z)$ of the jerk models by the use of
OHD+H$_{2.3}$. The original data points are also shown in the same
plane for comparison. The importance of H$_{2.3}$ is also clear
because the whole point (with error bar) locates in the 1$\sigma$
errors of $H(z)$.

As another important parameter, the equation of state $\omega$ of
the ``dark energy'' plays a crucial role in explaining the cosmic
acceleration. On the other hand, $j(z)$ can be also interpreted in
fluid term through the relation\cite{j..Blandford}
\begin{equation}
  j=-1+\frac{4\pi \dot{P}}{H^{3}},
\end{equation}
where $P$ is the pressure and the gravitational constant $G$ is set to be unity. This expression is valid without
considering the spatial curvature. Moreover, this equation derives
the relation between $j$ and $w$ as in Ref.\cite{j..Blandford} where
a constant $w$ is assumed. It is also interesting to calculate $w$
in our jerk parameterizations. If we assume that matter evolves in
the usual way, the first term in the expression of $E(z)$ in
Eq.(\ref{Eq:E1})-(\ref{Eq:E4}) can be treated as the matter term
with $ C_{1}/3=\Omega_{m0}$, while the rests stand for the ``dark
energy'' term. Specifically, the models can be summarized as
\begin{equation}
  E^{2}(z)=\Omega_{m0}(1+z)^{3}+\Omega_{DE}(z),
\end{equation}
where the subscript ``DE'' represents the dark energy term. On the
other hand, from the Friedmann equation, we can obtain $E(z)$ in a
universe comprised by the matter and exotic dark energy
\begin{equation}
  E^{2}(z)=\Omega_{m0}(1+z)^{3}+\Omega_{0}\exp\left[3\int_{0}^{z}\frac{1+\omega(z')}{1+z'}dz'\right].
\end{equation}
The equivalence of these two expressions can give a relation between
$\omega(z)$ and $j(z)$ indirectly
\begin{equation}
  \omega(z)=\frac{\Omega_{DE}'(z)}{3\Omega_{DE}(z)}(1+z)-1.
\end{equation}
In the right pannel of Fig.\ref{fig:recon_H}, we plot $w(z)$ of the
four jerk models. The best-fit results show $\omega(z)<-1$ in the
past, and the departure from -1 enlarges as the redshift $z$
increases. More importantly, the current value $\omega_{0}$ is not
strict -1 but a little smaller. This does not contradict with
$j(z=0)=-1$ in our calculation since both $j$ and $\omega$ are not
constant. It is different from the previous work \cite{j..Blandford}
where a constant $\omega$ is assumed. Taking into account the
uncertainty, the $\Lambda$CDM model can be well accommodated in the
1$\sigma$ errors.

\subsection{The deceleration factor $q(z)$ and $Om(z)$ diagnostic}

\begin{figure*}
%\vspace{.2in}
\centerline{\psfig{figure=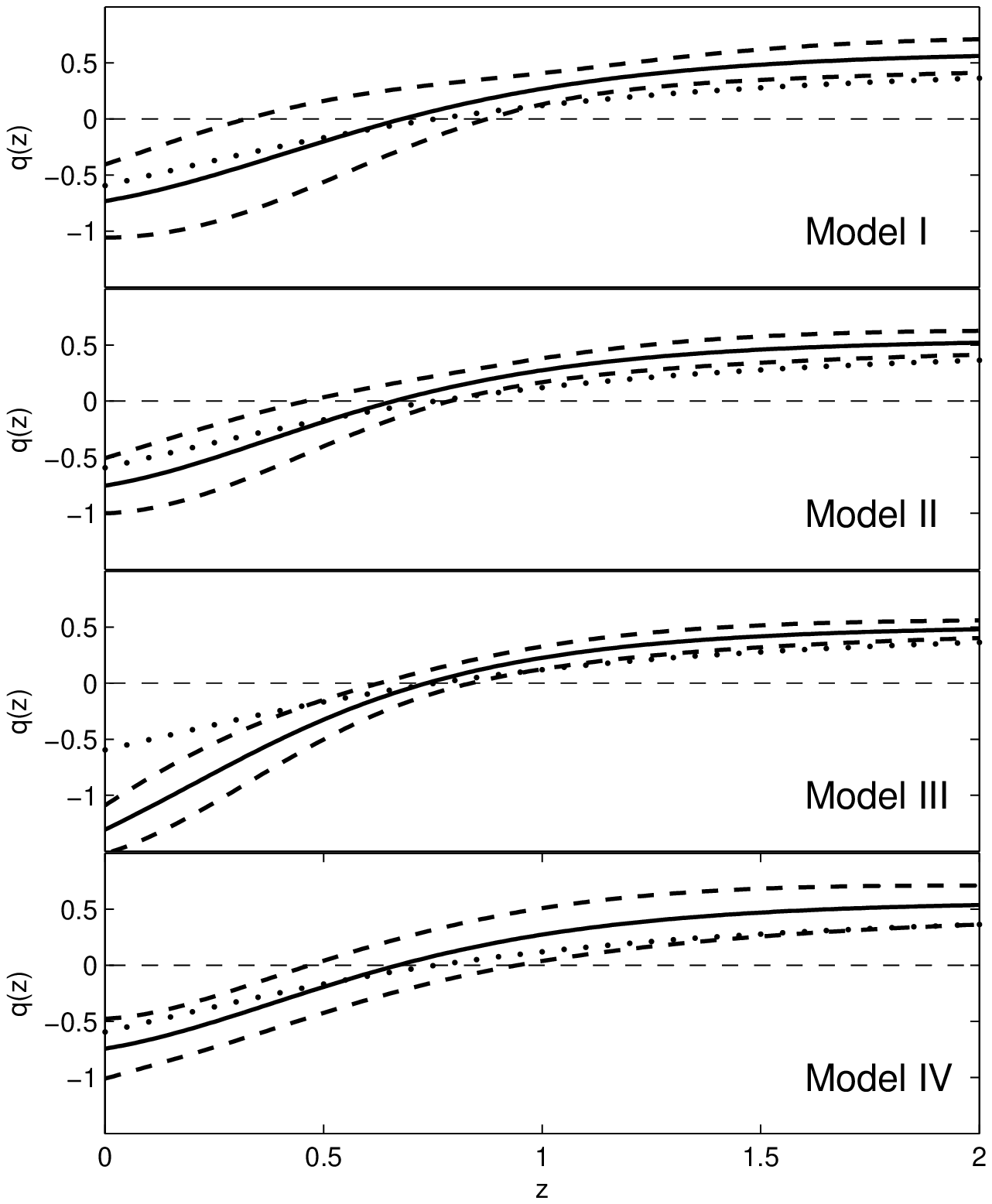,width=3.5truein,height=6.5truein}
\psfig{figure=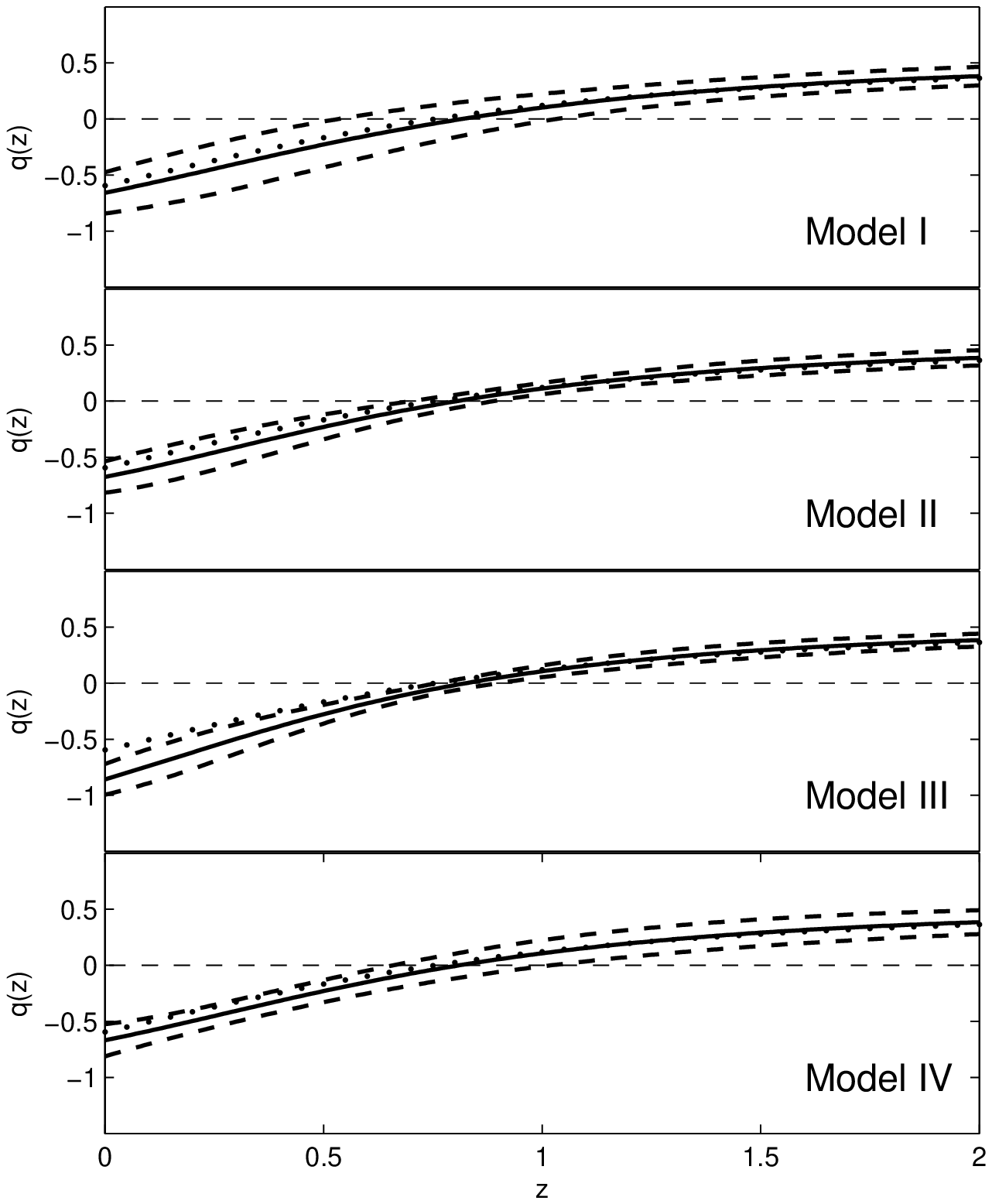,width=3.5truein,height=6.5truein}
\hskip 0.01in} \caption{The reconstruction of the deceleration
factor. The solid lines represent the best-fit curves, while the
dashed lines stand for the $1\sigma$ error. $Left$: OHD; $Right$:
OHD+H$_{2.3}$. The dotted lines stand for the spatially flat
$\Lambda$CDM model with $\Omega_{m0}=0.27$ \cite{CMB..Komatsu} and
the horizonal thin dashed lines represent $q=0$.
}\label{fig:recon_q}
\end{figure*}

As a further step in tracing the dynamic of the jerk
parameterizations, we also calculate the deceleration factor $q(z)$
as Eq.(\ref{Eq:q}) shows. Our results are presented in
Fig.\ref{fig:recon_q}. Since the SNe data sample supports the
standard $\Lambda$CDM model, we do not plot its curve and $1\sigma$
errors and only plot the ones of OHD. We use $q(z)$ of the
$\Lambda$CDM to represent the SNe results. We find that OHD
observations also show an accelerated expansion of the universe in
our jerk parameterizations. Despite the bad constraint of the
parameters, this phenomenon is proved at high confidence level. One
may attribute the reason of the accelerated expansion to the nearly
zero value of $j_{1}$ because $j_{1}=0$ leads to $j=-1$ and returns
to $\Lambda$CDM model. However, this should be caused by the value
of $C_{1}$ in the jerk parameterizations, because a matter-dominated
universe (no accelerated expansion) also has $j=-1$ as we discussed
in the preceding section.

In addition, $q(z)$ of Model III in Fig.\ref{fig:recon_q} behaves
differently from other models. The current value $q(z=0)$ is not
consistent with $\Lambda$CDM model which can be seen as a deviation
at certain extent (even the addition of H$_{2.3}$ does not change
this). This is not unexpected as seen in Fig.\ref{fig:recon_j}, the
reconstruction of $j(z)$ in Model III. The slope of the best-fit
curve at $z=0$ is the largest in Model III which shows a remarkable
deviation from $j(z)=-1$ among these models. But we have to say this results
are obtained by the use of OHD sample with the first $H_{0}$ prior. When the
second prior is used, the deviation disappeared. This is because the constraints
are affected at some extent by the choice of prior \cite{OHD..Farooq}.

Another important epoch of the evolution of the universe is the
transition redshift $z_{t}$ when the universe began to expand with
an acceleration from the cosmic deceleration phase
\cite{OHD..Busca,transition..Lima,transition..Farooq1,transition..Farooq2}.
Their works studied the possibility of using the transition redshift as a potential
cosmic variable. Except that, their results also show the prediction as the standard
$\Lambda$CDM indicates.
In the present work, the results obtained from OHD in
Fig.\ref{fig:recon_q} show that $0.4<z_{t}<1$ for the four jerk
parameterizations. This is consistent with the previous works but
the uncertainty is obvious \cite{j..model2..2}. The addition of
H$_{2.3}$ improves this estimation as expected, and the consistency
between the jerk models may lead us to believe that the transition
redshift is not an artifact of the parameterizations.

In our reconstruction, the second term in Eq.(\ref{Eq:re_j}) can be
treated as a perturbation from the standard $\Lambda$CDM model and
$j_{1}$ is a measurement of its magnitude. In addition, $j_{1}$
works differently in the four parameterizations. This can be found
from the results of $H(z)$, $\omega(z)$ and $q(z)$. Although the
calculations show that the different parameterizations have
consistent results with each other, it is still of some necessity to
find the influence of different functions in choosing $f(z)$ in
Eq.(\ref{Eq:re_j}).

As we know, except for the statefinder diagnostic, another common
tool in distinguishing dark energy models is the $Om(z)$ diagnostic
which is defined as \cite{Om..Sahni}
\begin{equation}
  Om(z)=\frac{E^{2}(z)-1}{(1+z)^{3}-1}.
\end{equation}
Apparently, $Om(z)=\Omega_{m0}$ for $\Lambda$CDM model, therefore
this function is useful and powerful in distinguishing $\Lambda$CDM
from other dark energy models. Additionally, since $Om(z)$ relies
only on the knowledge of Hubble parameter or equivalently, the
expansion factor, the errors in the reconstruction of $Om$ are bound
to be small.

On the other hand, since $Om(z)=\Omega_{m0}$ which is the fraction
of the matter term in the $\Lambda$CDM model, one may conjecture
that the deviation of $Om(z)$ from a constant value $\Omega_{m0}$ in
certain dark energy models can represent a perturbation that comes
from the effect rather than the matter.

\begin{figure*}
%\vspace{.2in}
\centerline{\psfig{figure=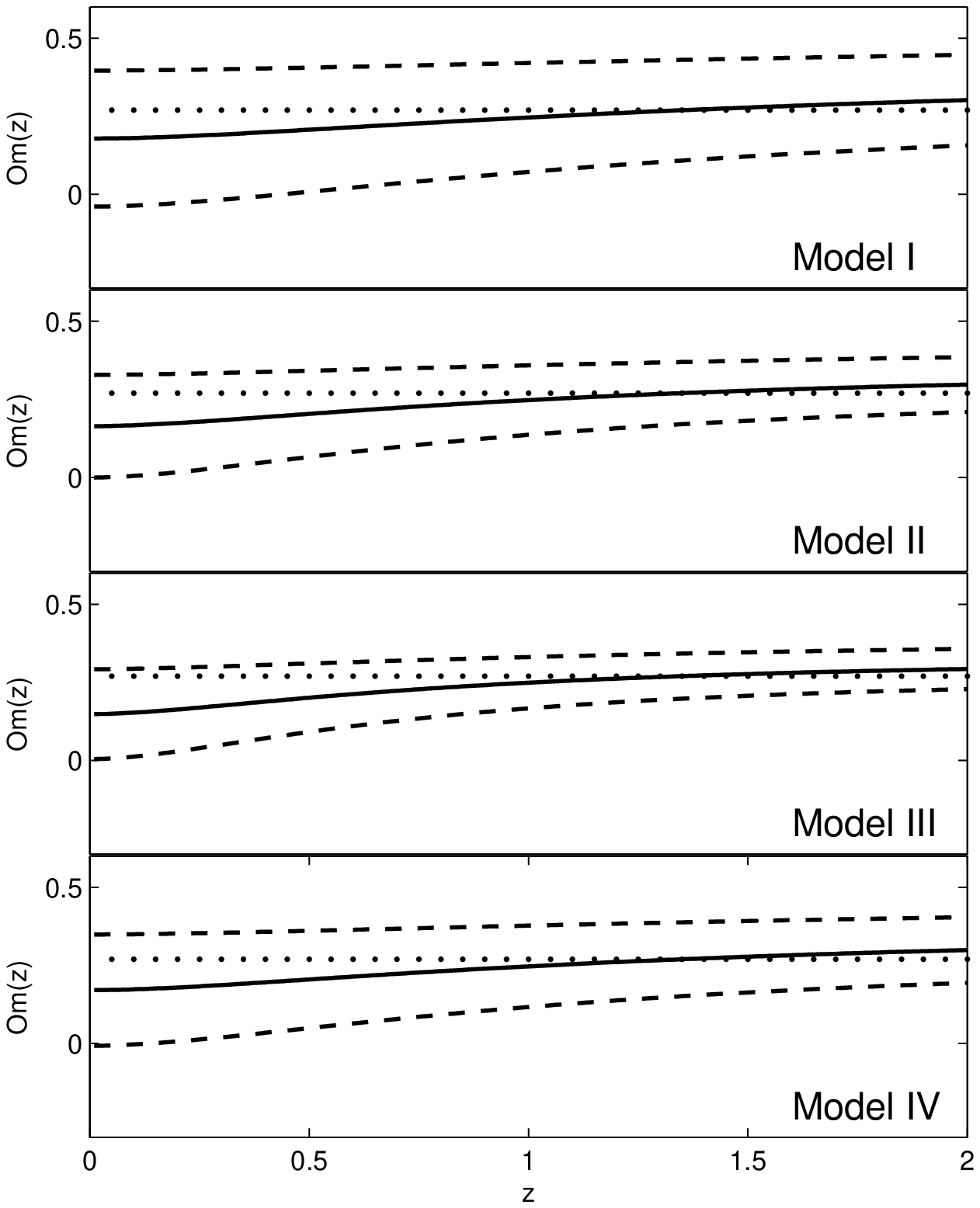,width=3.5truein,height=6.5truein}
\psfig{figure=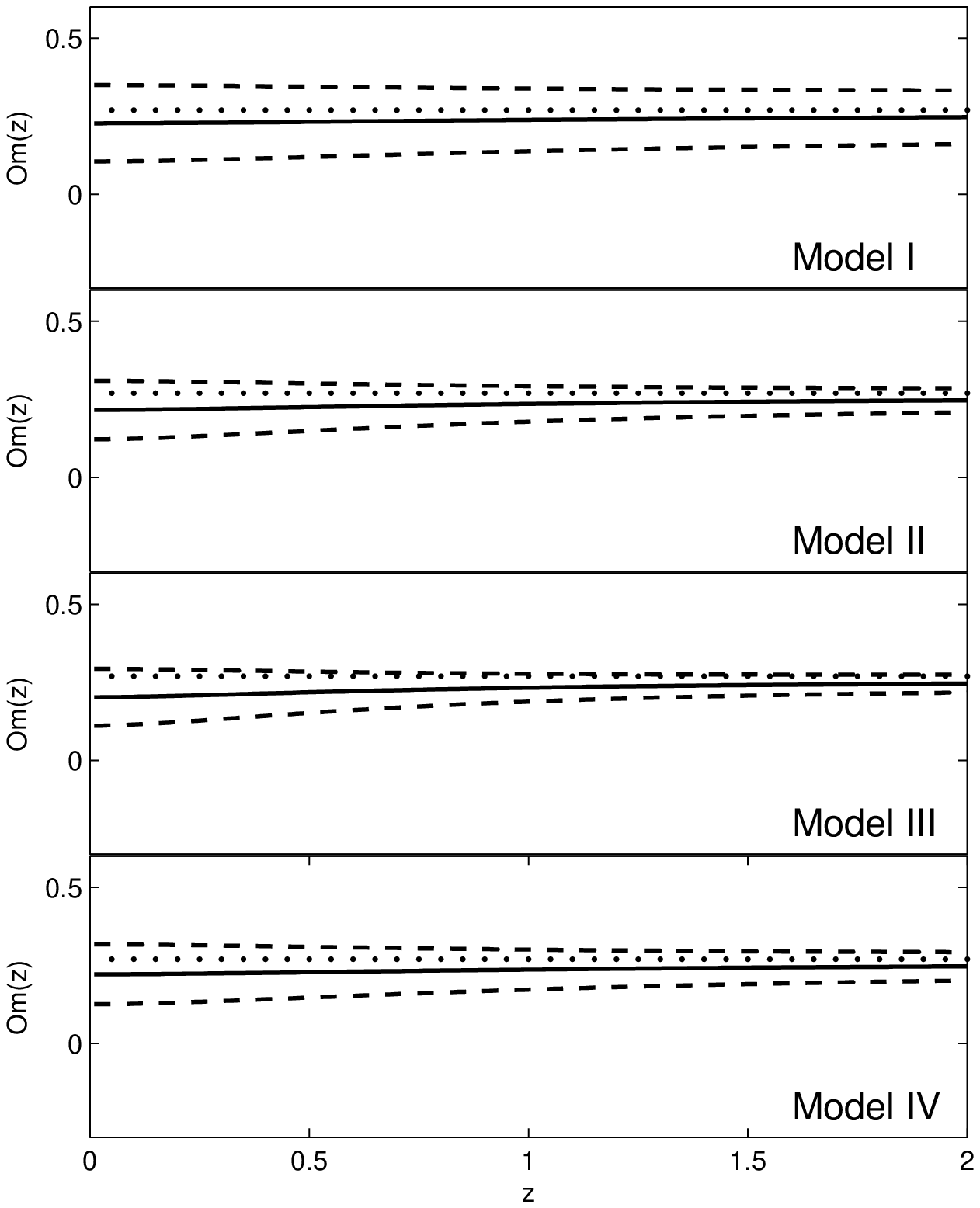,width=3.5truein,height=6.5truein}
\hskip 0.01in} \caption{The reconstruction of the $Om(z)$
diagnostic. The solid lines represent the best-fit curves, while the
dashed lines stand for the $1\sigma$ error. $Left$: OHD; $Right$:
OHD+H$_{2.3}$. The dotted lines stand for the spatially flat
$\Lambda$CDM model with $\Omega_{m0}=0.27$. }\label{fig:recon_Om}
\end{figure*}

The reconstruction results of $Om(z)$ are presented in
Fig.\ref{fig:recon_Om}. The SNe constraint results are just
represented by the standard $\Lambda$CDM curve and the $1\sigma$
error is neglected. We can find that the $\Lambda$CDM model can be
well accommodated by OHD and OHD+H$_{2.3}$ samples, but the best-fit
values favor a smaller matter proportion in the low redshift range.
Except that, the reconstructed evolution curves of all the four jerk
models change little in the redshift range $0<z<2$, especially for
the OHD+H$_{2.3}$ one. This is caused by the relatively small value
of $j_{1}$ and this can be seen as a proof of treating the $j_{1}$
term in Eq.(\ref{Eq:re_j}) as a perturbation.

%===================figure4=======================
%=================================================
%===================figure5=======================
%=================================================

\subsection{The joint constraints}

%===================figure2_1=======================
\begin{figure*}
 \center
\includegraphics[scale=0.6]{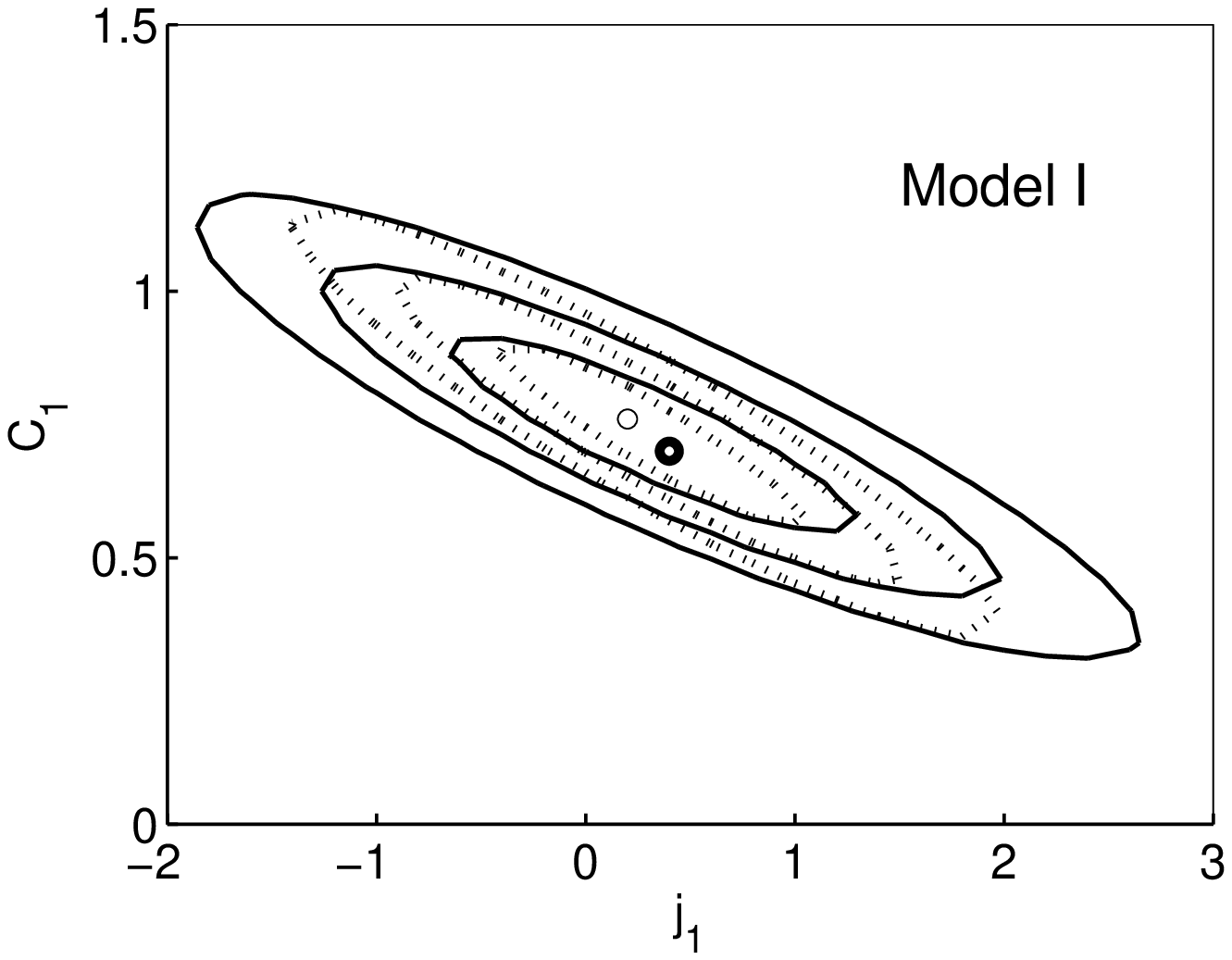}
\includegraphics[scale=0.6]{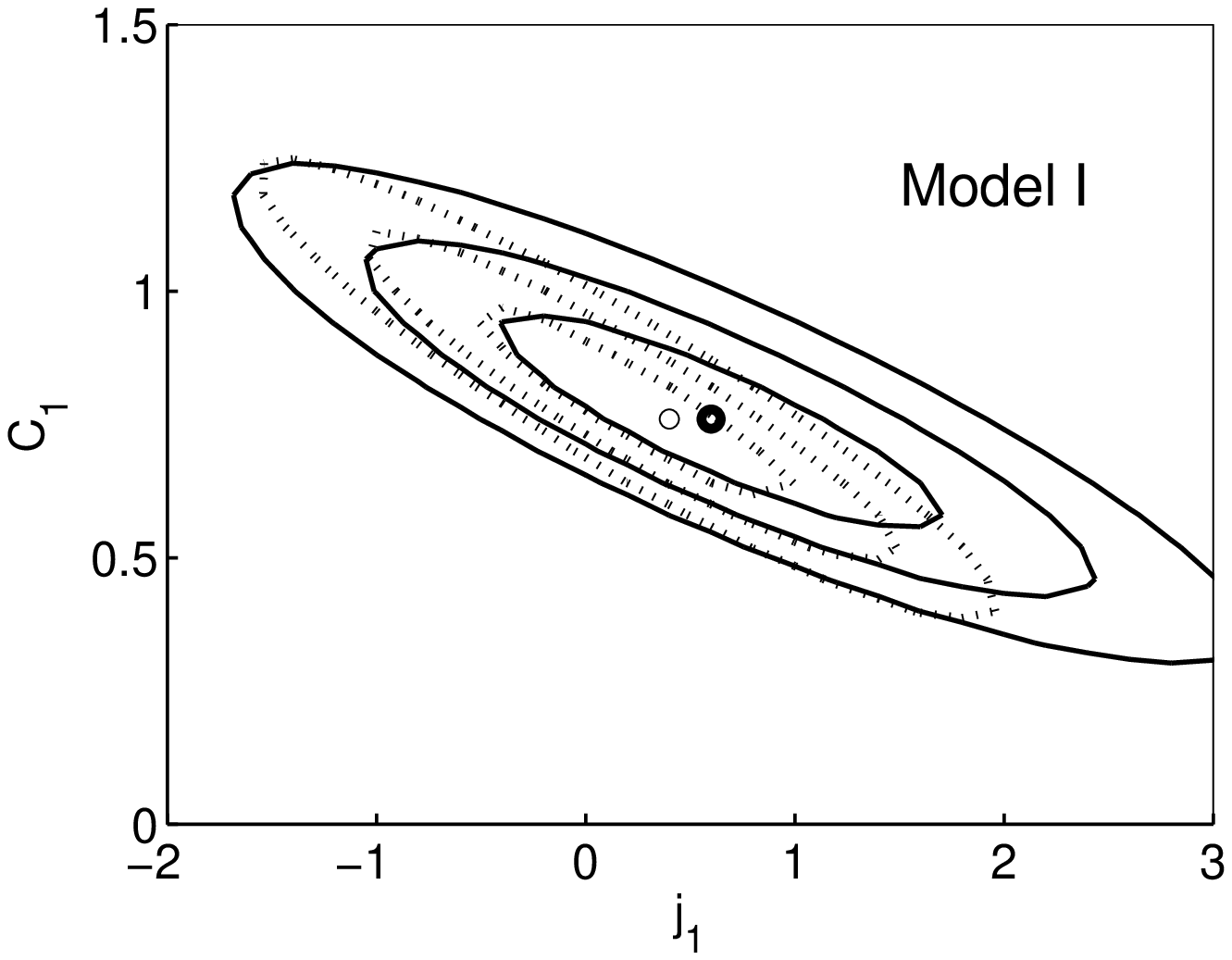}
\caption{The joint constraints for Model I. $Left$: the OHD with prior of $H_{0}=74.3\pm2.1$ km s$^{-1}$ Mpc$^{-1}$ and
SNe data with (solid) and without (dotted) systematic errors. $Right$: the OHD with prior of $H_{0}=68\pm2.8$ km s$^{-1}$ Mpc$^{-1}$ and
SNe data with (solid) and without (dotted) systematic errors. The thin and thick circles belong to dotted and solid contours respectively.}
\label{fig:joint_1}
\end{figure*}
%=================================================
%===================figure2_2=======================
\begin{figure*}
 \center
\includegraphics[scale=0.6]{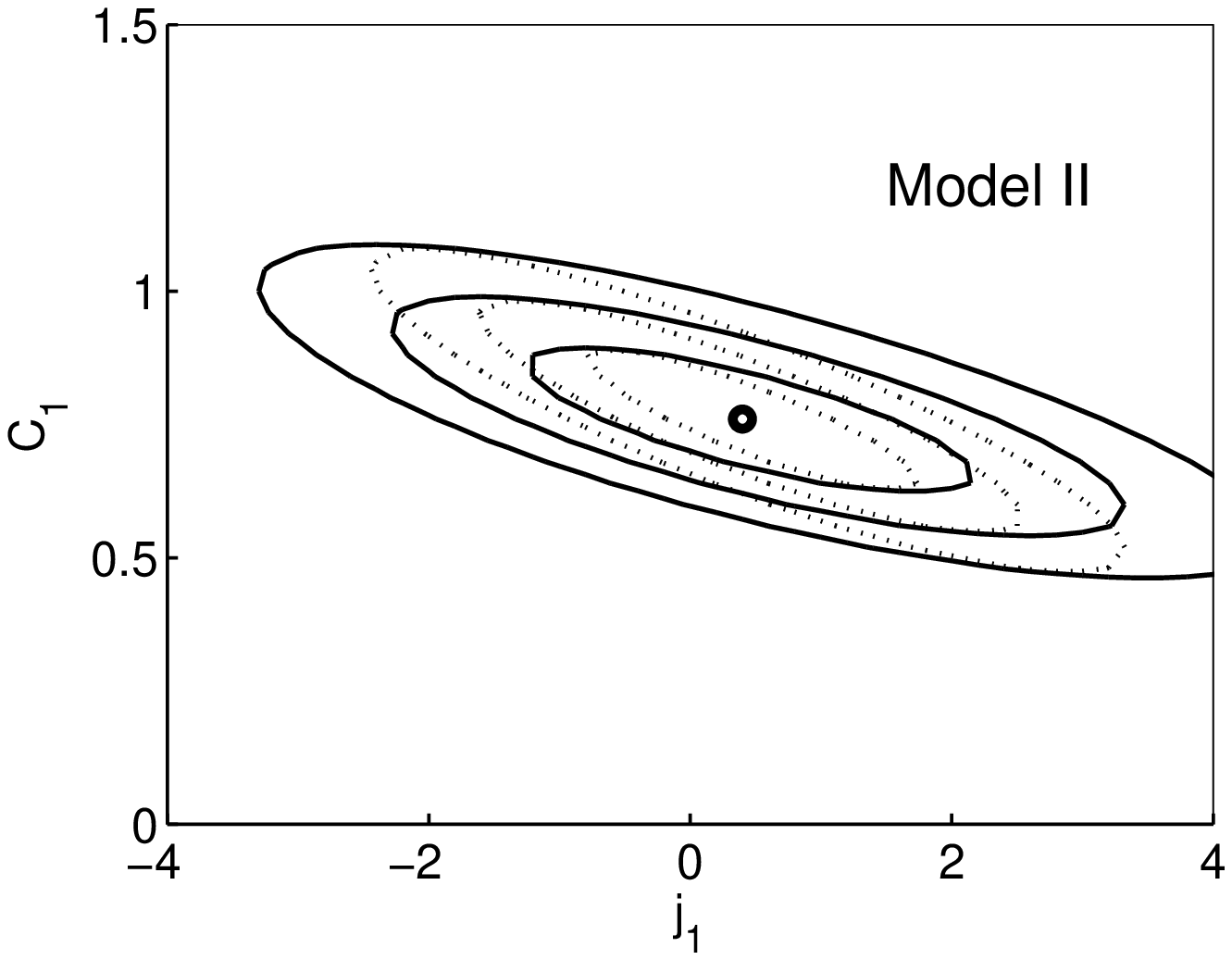}
\includegraphics[scale=0.6]{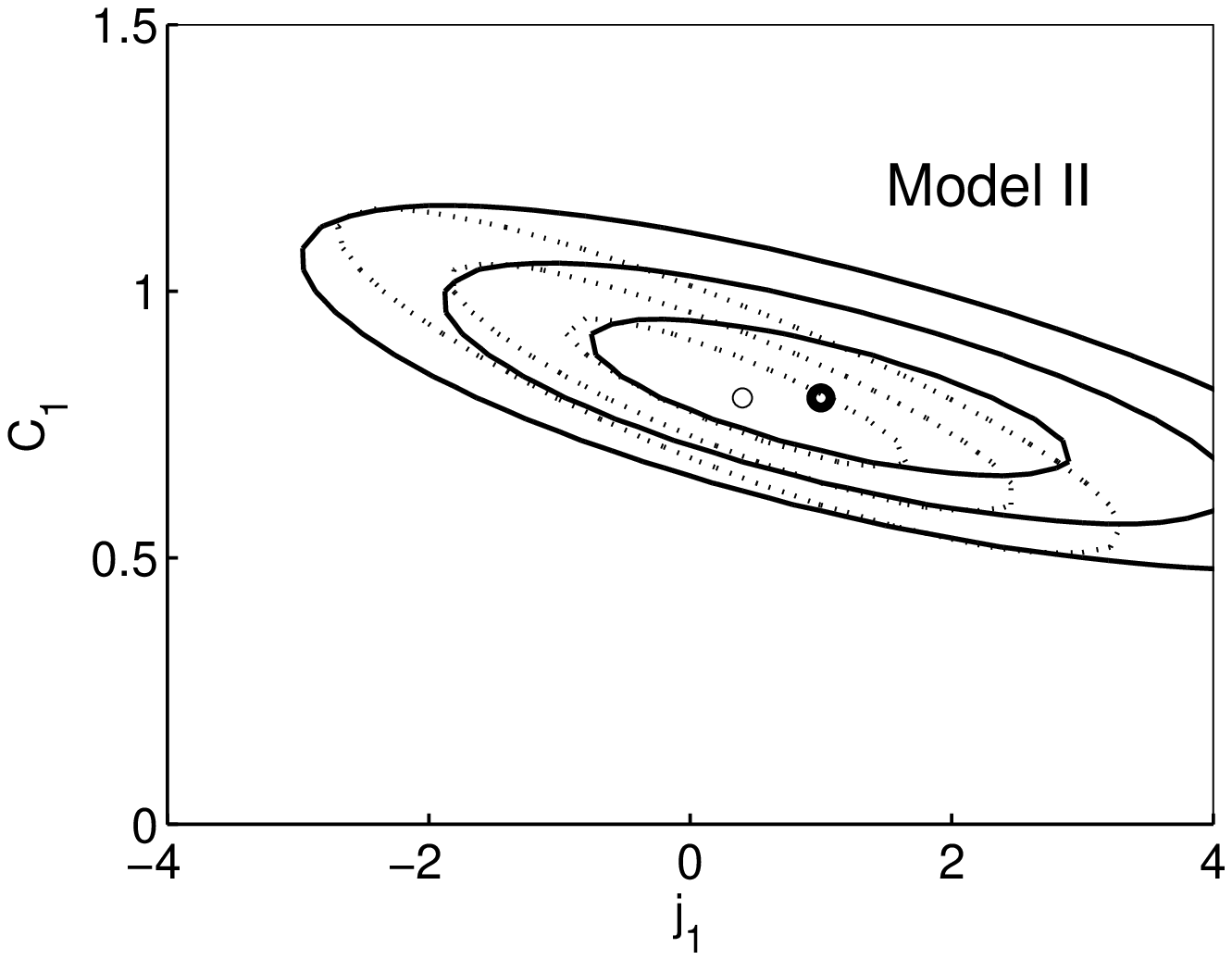}
\caption{The same as FIG.\ref{fig:joint_1} but for Model II.} \label{fig:joint_2}
\end{figure*}
%=================================================
%===================figure2_3=======================
\begin{figure*}
 \center
\includegraphics[scale=0.6]{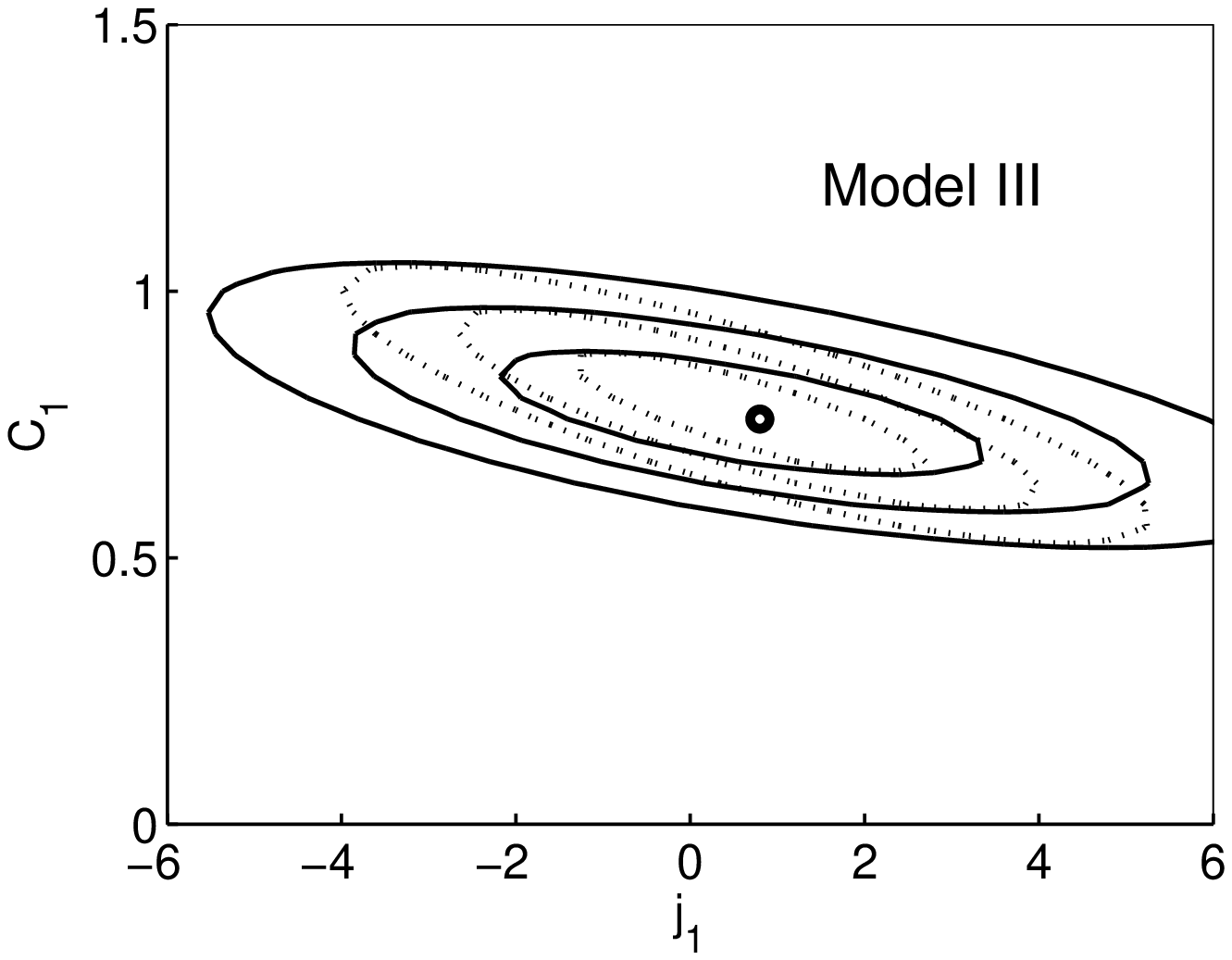}
\includegraphics[scale=0.6]{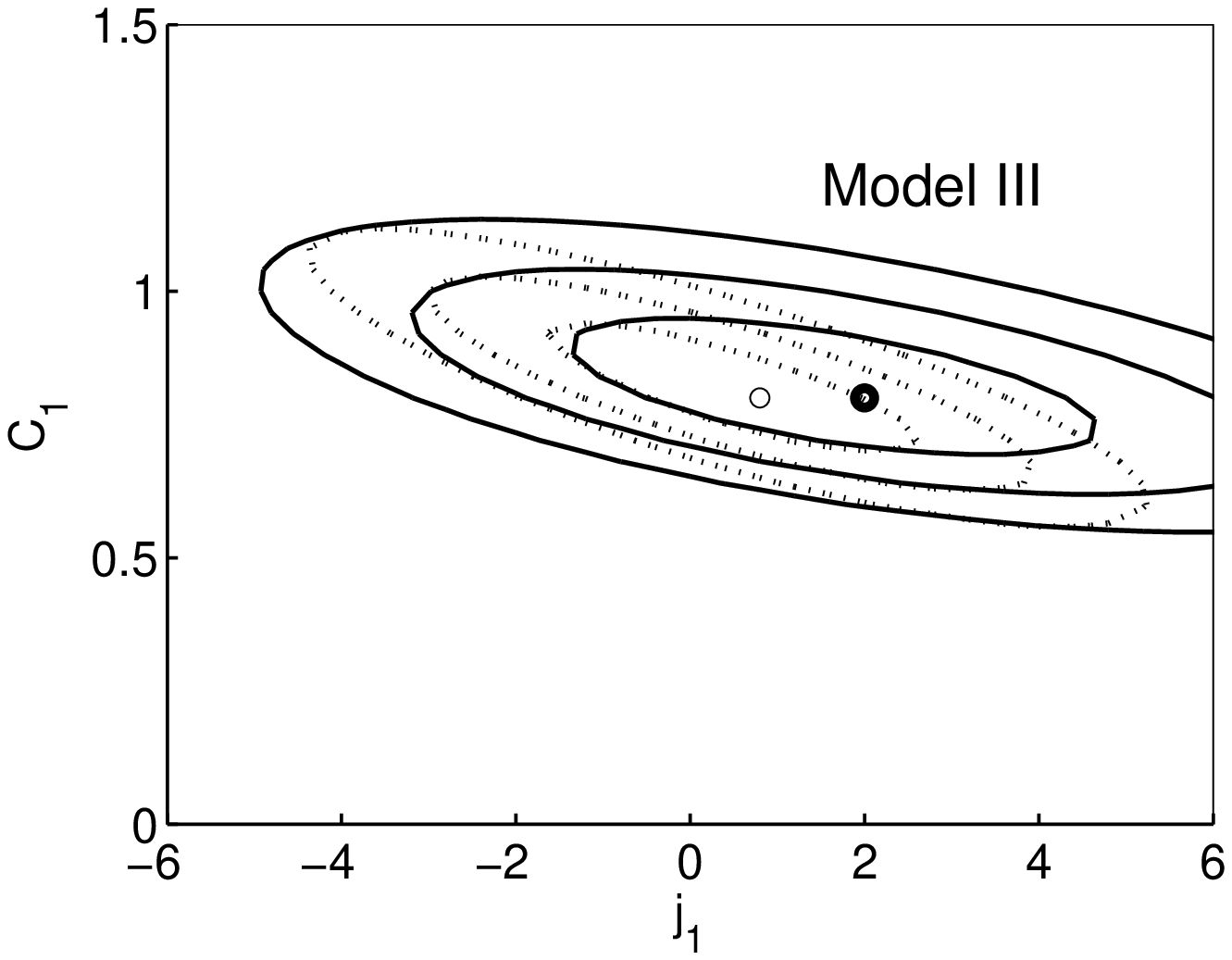}
\caption{The same as FIG.\ref{fig:joint_1} but for Model III.} \label{fig:joint_3}
\end{figure*}
%=================================================
%===================figure2_4=======================
\begin{figure*}
 \center
\includegraphics[scale=0.6]{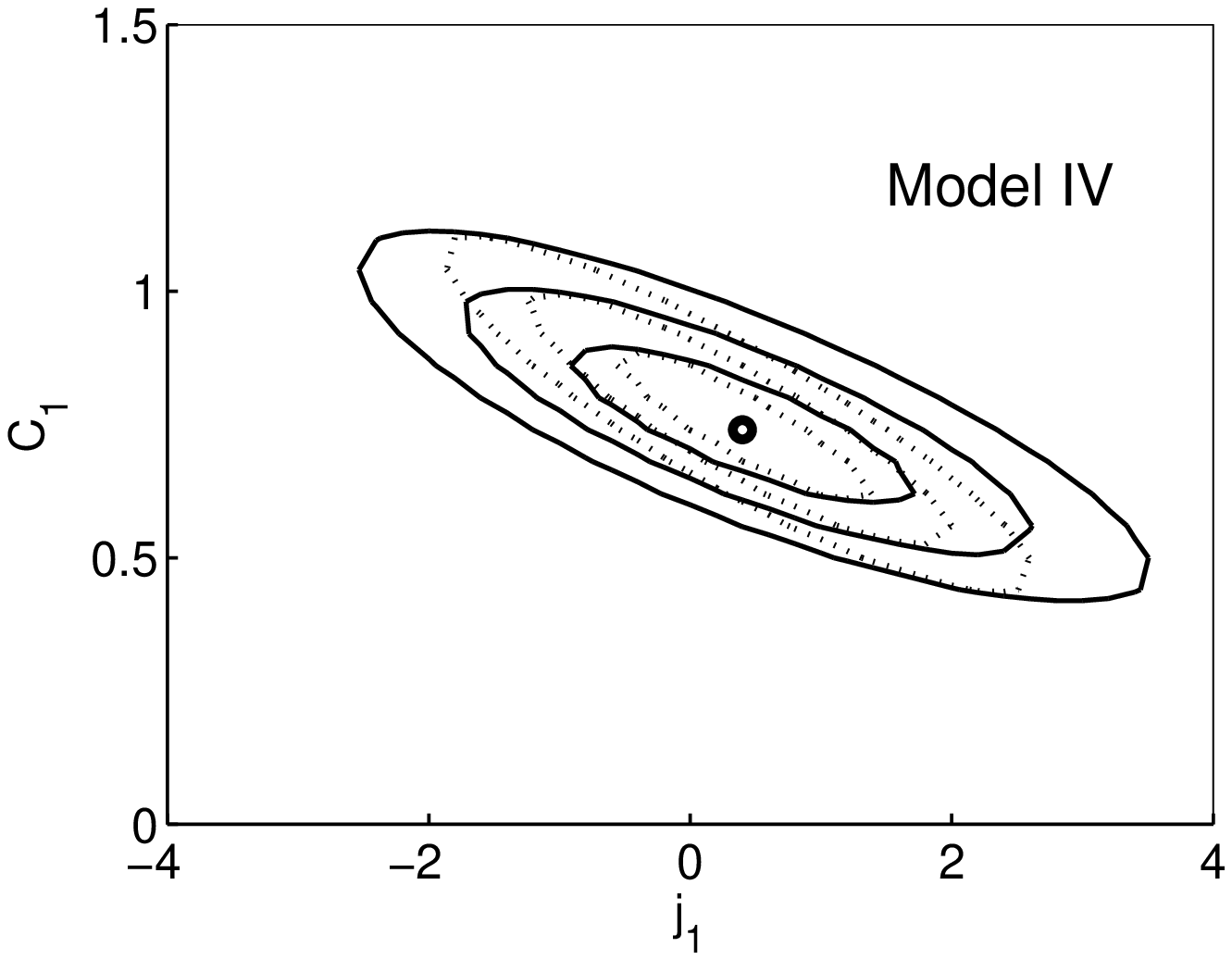}
\includegraphics[scale=0.6]{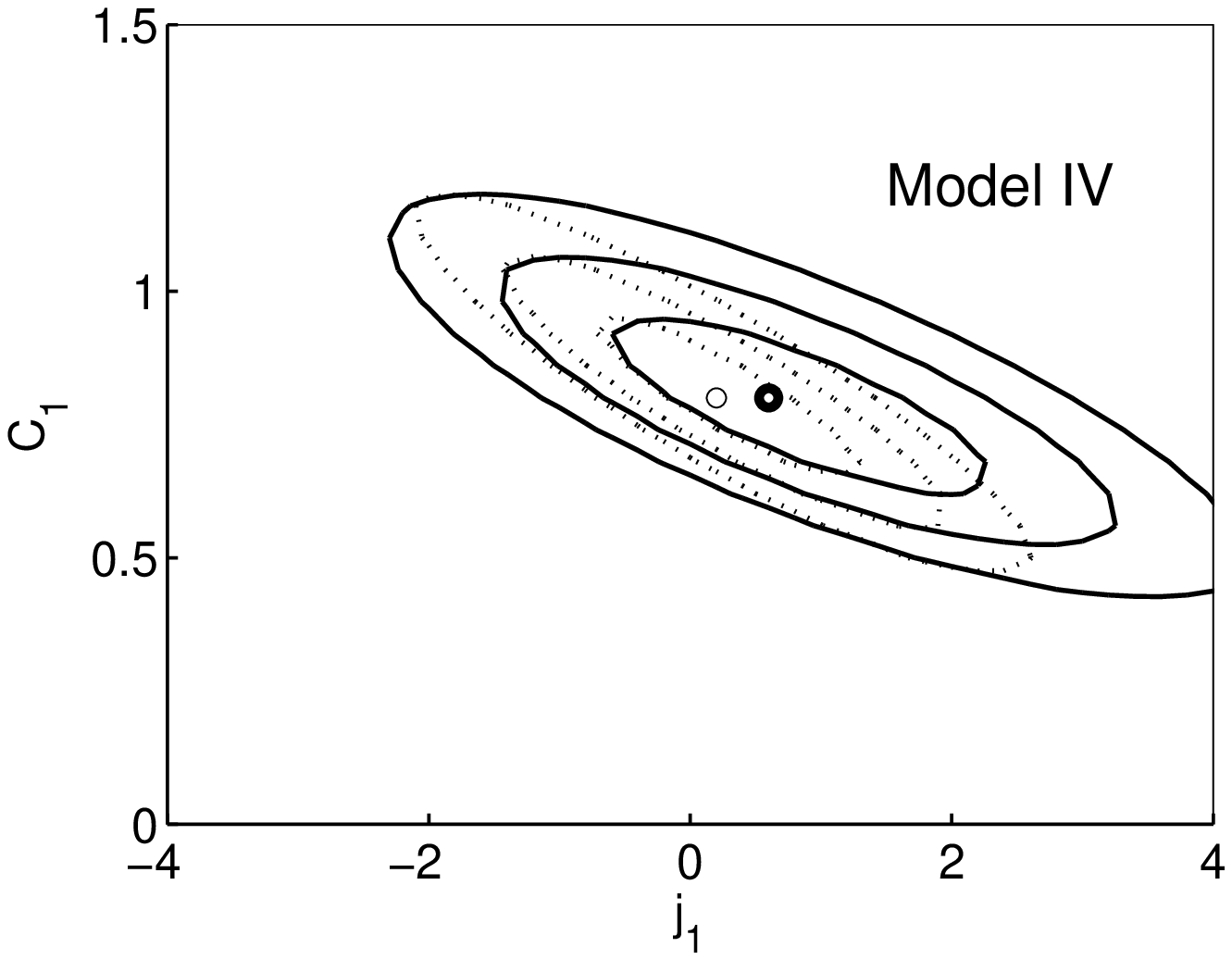}
\caption{The same as FIG.\ref{fig:joint_1} but for Model IV.} \label{fig:joint_4}
\end{figure*}
%=================================================

From the previous results, we can see that the SNe and OHD samples give similar constraints
on the jerk parameterizations models. Therefore one may expect that the joint constraint of these
data could provide tighter constraints. This will give more accurate values of the cosmological parameters
and help to discriminate different cosmological models. Our results of the joint constraints are presented
in FIG.\ref{fig:joint_1} to FIG.\ref{fig:joint_4} and TABLE.\ref{tab:results} where different combinations of the data samples SNe (with and without systematic errors)
and OHD (two choices of the prior of $H_{0}$) are considered. The best-fit values of the parameters are confirmed by minimizing
\begin{equation}
  \chi^{2}=\chi_{SNe}^{2}+\chi_{OHD}^{2}.
\end{equation}
And the confidence regions are found by the same method as the single constraints.

From the values of FoM, we can see that the joint data improve the constraints significantly. And the best
constraint comes from SNe without considering systematic errors plus OHD+H$_{2.3}$ with the prior of $H_{0}=74.3\pm2.1$ km s$^{-1}$ Mpc$^{-1}$.
From the confidence regions, we can conclude that the second prior of $H_{0}$ give worse constraints than the first one.
However, it also favors a larger value of $j_{1}$.

The joint constraints give consistent results of the parameter $C_{1}$ and their uncertainties are also sufficiently small.
Together with the fact of small value of $j_{1}$, this shows the proportion of the matter in the universe.

\begin{table*}
\begin{center}\begin{tabular}{l||c|c|c|c|c}
 %\MC{3}{c}{\text{caption}}\\[5pt]
 \hline
               &      data        &        $j_{1}$        &         $C_{1}$         &      $\chi_{\text{min}}^{2}$       &      FoM                       \\ \hline
 %\MC{3}{|c|c|}{\ZZ{-8pt}{15pt}\hfill\normalsize   \hfill  \hfill\normalsize MGCDM     \hfill\normalsize $\Lambda$CDM  }\\ \hline
 %\ZZ{-6pt}{22pt}
 \normalsize Model I          &  SNe$_{a}$+OHD+H$_{2.3}$($H_{0}'$)              &  $0.20\pm0.56$          &   $0.76\pm0.13$   &   $584.82$    &  47.8868  \\ \hline
 \normalsize Model I          &  SNe$_{b}$+OHD+H$_{2.3}$($H_{0}'$)                  &  $0.40\pm0.59$          &   $0.70\pm0.10$   &   $569.16$    &  26.2562  \\ \hline
 \normalsize Model I          &  SNe$_{a}$+OHD+H$_{2.3}$($H_{0}$)              &  $0.40\pm0.39$          &   $0.76\pm0.09$   &   $582.99$    &  42.2382  \\ \hline
 \normalsize Model I          &  SNe$_{b}$+OHD+H$_{2.3}$($H_{0}$)                  &  $0.60\pm0.73$          &   $0.76\pm0.14$   &   $566.41$    &  21.3222  \\ \hline
 \normalsize Model II         &  SNe$_{a}$+OHD+H$_{2.3}$($H_{0}'$)                &  $0.40\pm0.87$          &   $0.76\pm0.09$   &   $584.95$    &  27.5121  \\ \hline
 \normalsize Model II         &  SNe$_{b}$+OHD+H$_{2.3}$($H_{0}'$)                  &  $0.40\pm1.14$        &   $0.76\pm0.09$   &   $569.18$    &  15.0537  \\ \hline
 \normalsize Model II         &  SNe$_{a}$+OHD+H$_{2.3}$($H_{0}$)                &  $0.40\pm0.81$          &   $0.80\pm0.08$   &   $582.88$    &  24.7434  \\ \hline
 \normalsize Model II         &  SNe$_{b}$+OHD+H$_{2.3}$($H_{0}$)                  &  $1.00\pm1.25$        &   $0.80\pm0.10$   &   $566.50$    &  12.1524  \\ \hline
 \normalsize Model III        &  SNe$_{a}$+OHD+H$_{2.3}$($H_{0}'$)               &  $0.80\pm1.27$          &   $0.76\pm0.07$   &   $584.91$    &  17.1361  \\ \hline
 \normalsize Model III        &  SNe$_{b}$+OHD+H$_{2.3}$($H_{0}'$)                  &  $0.80\pm1.69$       &   $0.76\pm0.07$   &   $569.25$    &  9.2012  \\ \hline
 \normalsize Model III        &  SNe$_{a}$+OHD+H$_{2.3}$($H_{0}$)               &  $0.80\pm1.19$          &   $0.80\pm0.07$   &   $582.96$    &  15.3857  \\ \hline
 \normalsize Model III        &  SNe$_{b}$+OHD+H$_{2.3}$($H_{0}$)                  &  $2.00\pm1.74$       &   $0.80\pm0.07$   &   $566.64$    &  7.3906  \\ \hline
 \normalsize Model IV         &  SNe$_{a}$+OHD+H$_{2.3}$($H_{0}'$)               &  $0.40\pm0.64$          &   $0.74\pm0.08$   &   $584.81$    &  36.1403  \\ \hline
 \normalsize Model IV         &  SNe$_{b}$+OHD+H$_{2.3}$($H_{0}'$)                  &  $0.40\pm0.87$        &   $0.74\pm0.09$   &   $569.15$    &  19.6640  \\ \hline
 \normalsize Model IV         &  SNe$_{a}$+OHD+H$_{2.3}$($H_{0}$)               &  $0.20\pm0.73$          &   $0.80\pm0.09$   &   $582.94$    &  31.9332  \\ \hline
 \normalsize Model IV         &  SNe$_{b}$+OHD+H$_{2.3}$($H_{0}$)                  &  $0.60\pm1.10$        &   $0.80\pm0.12$   &   $566.52$    &  15.6984  \\ \hline
\end{tabular}
\end{center}
\caption{The constraint results of the parameters, including the
best-fit values with $1\sigma$ errors of the parameters and the FoM
of four jerk parameterizations. (The subscripts "$a$","$b$" and prime have the same meanings as TABLE.\ref{tab:SNe} and
TABLE.\ref{tab:OHD}). }\label{tab:results}
\end{table*}
\section{discussions and conclusion}\label{conclusion}

In this paper, we study the property of the cosmological jerk
parameter in detail. Just within the assumption of a homogeneous and
isotropic universe without any introduction of the underlying
gravitational theory and energy components, we propose several
various kinds of reconstruction of jerk parameter and perform a
kinematic analysis and constraint by the SNe and OHD samples.

Our constraining results show that the standard $\Lambda$CDM model
can be well accommodated by the SNe and OHD observations. In other
words, the $\Lambda$CDM model can be well supported by the evolution
of the scale factor up to its third time derivative. Especially, the
constraint of SNe data gives a nearly zero value of the parameter
$j_{1}$ which is a measurement of the deviation of $j(z)$ from the
$\Lambda$CDM model. Once the constraint uncertainties are taken into
account, the OHD can give similar results. This is consistent with
the previous studies that OHD can play the same role as SNe in
constraining cosmological models.

As pointed out in Sec.\ref{sec:re_j}, the parameterizations of jerk
in our work have $j=-1$ at $z=0$. Although the previous works did
not show the current universe is strictly the $\Lambda$CDM
one\cite{j..model2..1,j..model2..2,j..Rapetti}, the concordance
$\Lambda$CDM paradigm is also strongly supported \cite{j..Rapetti}.
Moreover, the real value of $j$ at $z=0$ should be directly obtained
by the current or at least, the low-redshift observations. However,
the low-redshift research is far from comprehensive or efficient
enough. For example, the equation of state of the dark energy is
allowable and theoretically possible to have arbitrarily large
fluctuations at ultra-low redshifts \cite{eos..Mortonson}.
This may increase the uncertainties of the measurements at low redshift.
And thus the study of the low redshift cosmology is
necessary and of great importance. Except that, There also appears
to be some tension between low $z$ and high $z$ data
\cite{tension..Shafieloo,tension..Wei}. These facts make the
confirmation of the current value of $j$ difficult. Thus compared
with this potential uncertainty, one possible way is to adopt the
simplest model, the $\Lambda$CDM model and ignore the chaotic
knowledge.

In order to obtain more information of our reconstruction, we also
calculated the Hubble parameter $H(z)$, equation of state $\omega$,
deceleration factor $q(z)$ and $Om(z)$ diagnostic of the jerk
models. The effect of the addition of H$_{2.3}$ is significant in
the $\chi^{2}$ constraint which can be seen from
Fig.\ref{fig:recon_H}. Also, the equation of state $\omega<-1$ in
the redshift range $0<z<2$ shows a phantom-like universe but the
evidence is not strong. The reconstruction of $q(z)$ indicate a
current accelerated expansion phase following a matter-dominated
phase. This is consistent with the great discovery of Type Ia SNe
\cite{SNe..Riess,SNe..Perlmutter}. Moreover, we employ the $Om(z)$
diagnostic in our calculation. This kind of function can not only
behave as a diagnostic to distinguish different dark energy models
from $\Lambda$CDM model. Moreover, We conjecture that it is also
useful in studying the kinematic models. Because this kind of models
is often obtained mathematically rather than physically, the
meanings of the parameters rising from the solution process are hard
to handle. From this point of view, $Om$ can be used as a ``matter
generator'' or the ``effective'' matter term in this kind of
cosmological models. Our results show that OHD prefers a smaller
value of $Om$ at $z=0$. Additionally, the change of $Om(z)$ in the
past was not significant. In other words, $Om(z)$ evolves like a
constant. This can be seen as a positive proof of regarding the
$j_{1}$ term in the $j(z)$ parameterizations as a perturbation.

In this work, we use two kinds of observational data: Type Ia
Supernovae and Hubble parameter. The comparisons between them in
constraining cosmological models have been studied for several
years. Because of the large size of the data sample and relatively
clear systematic errors, SNe always provides more efficient
constraint than OHD.
However, when the systematic errors of the measurements are taken into
account, the constraints from SNe become worse than OHD and this can be
seen in our jerk results.
Moreover, the latest development in the
measurement of Hubble parameter gives us the possibility that the
data sample with smaller size also has the power to constrain dark
energy models effectively even better than SNe \cite{OHD..Farooq2}.
Therefore, it is anticipated that the future high-z, high-accuracy
$H(z)$ determinations will provide more important contributions in
cosmological researches \cite{OHD..ZhangT}.

In our analysis, the joint constraints are also achieved. As expected, the joint constraints
can sufficiently improve the constraining results. And the different combinations
of the data samples also give us the information that the parameter $C_{1}$ is confirmed
in a high precision.

Additionally, the relationships of SNe and OHD in constraining jerk
model should be noticed. As we know, the measurement of SNe comes
from the distance modulus, while OHD comes from the ages of
passively evolving galaxies, BAO measurement and so forth. If we
regard the SNe observations as the distance measurement, OHD should
be related to the velocity measurement which is a time derivative of
the distance. Thus the information of acceleration or higher order
derivative (such as jerk, snap etc.) can be obtained from distance
or velocity measurement respectively. One question is that what is
the difference when we use distance and velocity respectively to
study the acceleration or jerk? An important issue is the error
propagation. The number of derivative from distance to acceleration
or higher-order variables is bigger than that from velocity to them.
So this may increase the uncertainty in estimating the acceleration and jerk.
From this point of
view, the OHD should be a better tool in studying the evolution of
the universe, especially when we hope to find more accurate and
subtle information of the universe since this kind of information
can be well carried by the higher order derivative of the scale
factor.
\section{acknowledgement}
We are very grateful to the anonymous referee for his valuable
comments and suggestions that greatly improve this paper. The
authors would like to thank W-B Liu for helpful discussions and
suggestions. This work is supported by the Ministry of Science and
Technology National Basic Science program (project 973) under grant
No. 2012CB821804, the National Natural Science Foundation of China
(Grant Nos. 11173006, 10875012, 11235003 and 11365008), the Fundamental
Research Funds for the Central Universities and
the Natural Science Fund of Education department of Hubei Province (Grant No. Q20131901).

\end{document}